\documentclass[a4paper,11pt,reqno]{amsart}
\usepackage{latexsym, amsmath, amsfonts, amssymb, amsthm, amscd, epsfig}
\usepackage{url}

\usepackage[font=small, labelfont={sc}]{caption}
\usepackage{subfig}
\usepackage{makerobust}

\usepackage[a4paper,scale={0.73,0.75},marginratio={1:1},footskip=10mm,headsep=10mm]{geometry}

\usepackage{color}

\usepackage{hyperref}

\numberwithin{equation}{section}

\def \beq {\begin{eqnarray}}
\def \eeq {\end{eqnarray}}
\def \beqn {\begin{eqnarray*}}
\def \eeqn {\end{eqnarray*}}

\newtheorem{theorem}{Theorem}
\newtheorem{itlemma}[theorem]{Lemma}
\newtheorem{itproposition}[theorem]{Proposition}
\newtheorem{itcorollary}[theorem]{Corollary}
\newtheorem{itremark}[theorem]{Remark}
\newtheorem{itdefinition}[theorem]{Definition}
\newtheorem{itexample}[theorem]{Example}
\newtheorem{itclaim}[theorem]{Claim}
\newtheorem{itfact}[theorem]{Fact}

\newenvironment{fact}{\begin{itfact}\rm}{\end{itfact}}
\newenvironment{claim}{\begin{itclaim}\rm}{\end{itclaim}}
\newenvironment{lemma}{\begin{itlemma}}{\end{itlemma}}
\newenvironment{remark}{\begin{itremark}\rm}{\end{itremark}}
\newenvironment{corollary}{\begin{itcorollary}}{\end{itcorollary}}
\newenvironment{proposition}{\begin{itproposition}}{\end{itproposition}}
\newenvironment{definition}{\begin{itdefinition}\rm}{\end{itdefinition}}
\newenvironment{example}{\begin{itexample}\rm}{\end{itexample}}

\newcommand{\be}[1]{\begin{equation}\label{#1}}
\newcommand{\ee}{\end{equation}}
\newcommand{\bl}[1]{\begin{lemma}\label{#1}}
\newcommand{\br}[1]{\begin{remark}\label{#1}}
\newcommand{\brs}[1]{\begin{remarks}\label{#1}}
\newcommand{\bt}[1]{\begin{theorem}\label{#1}}
\newcommand{\bd}[1]{\begin{definition}\label{#1}}
\newcommand{\bp}[1]{\begin{proposition}\label{#1}}
\newcommand{\bc}[1]{\begin{corollary}\label{#1}}
\newcommand{\bfact}[1]{\begin{fact}\label{#1}.}
\newcommand{\bex}[1]{\begin{example}\label{#1}.}
\newcommand{\ec}{\end{corollary}}
\newcommand{\efact}{\end{fact}}
\newcommand{\eex}{\end{example}}
\newcommand{\el}{\end{lemma}}
\newcommand{\er}{\end{remark}}
\newcommand{\ers}{\end{remarks}}
\newcommand{\et}{\end{theorem}}
\newcommand{\ed}{\end{definition}}
\newcommand{\ep}{\end{proposition}}
\newcommand{\epr}{\end{proof}}
\newcommand{\bpr}{\begin{proof}}
\newcommand{\bcl}[1]{\begin{claim}\label{#1}}
\newcommand{\ecl}{\end{claim}}

\newcommand{\ecs}{\end{corollary}}
\newcommand{\eers}{\end{exercise}}
\newcommand{\eexs}{\end{example}}
\newcommand{\eems}{\end{example}}
\newcommand{\els}{\end{lemma}}
\newcommand{\eles}{\end{lemmaex}}
\newcommand{\ets}{\end{theorem}}
\newcommand{\eds}{\end{definition}}
\newcommand{\eps}{\end{proposition}}

\newcommand{\bi}{\begin{itemize}}
\newcommand{\ei}{\end{itemize}}
\newcommand{\ben}{\begin{enumerate}}
\newcommand{\een}{\end{enumerate}}

\def\vbar{\mathchoice{\vrule height6.3ptdepth-.5ptwidth.8pt\kern-.8pt}
   {\vrule height6.3ptdepth-.5ptwidth.8pt\kern-.8pt}
   {\vrule height4.1ptdepth-.35ptwidth.6pt\kern-.6pt}
   {\vrule height3.1ptdepth-.25ptwidth.5pt\kern-.5pt}}
\def\fudge{\mathchoice{}{}{\mkern.5mu}{\mkern.8mu}}
\def\bbc#1#2{{\rm \mkern#2mu\vbar\mkern-#2mu#1}}
\def\bbb#1{{\rm I\mkern-3.5mu #1}}
\def\bba#1#2{{\rm #1\mkern-#2mu\fudge #1}}
\def\bb#1{{\count4=`#1 \advance\count4by-64 \ifcase\count4\or\bba A{11.5}\or
   \bbb B\or\bbc C{5}\or\bbb D\or\bbb E\or\bbb F \or\bbc G{5}\or\bbb H\or
   \bbb I\or\bbc J{3}\or\bbb K\or\bbb L \or\bbb M\or\bbb N\or\bbc O{5} \or
   \bbb P\or\bbc Q{5}\or\bbb R\or\bbc S{4.2}\or\bba T{10.5}\or\bbc U{5}\or
   \bba V{12}\or\bba W{16.5}\or\bba X{11}\or\bba Y{11.7}\or\bba Z{7.5}\fi}}

\def \Z {{\mathbb Z}}
\def \R {{\mathbb R}}

\def \N {{\mathbb N}}

\def \ra {\rightarrow }

\def \s {y}

\def \FF {{\cal{F}}}

\def \D{{\cal{D}}}

\def \ind {{\bf 1}}

\newcommand{\ba}[1]{\addtocounter{for}{1} \begin{eqnarray}\label{#1}}
\newcommand{\ea}{\end{eqnarray}}

\def\sqr#1#2{{\vcenter{\vbox{\hrule height .#2pt
                             \hbox{\vrule width .#2pt height#1pt \kern#1pt
                                   \vrule width .#2pt}
                             \hrule height .#2pt}}}}

\def\pmb#1{\setbox0=\hbox{#1}%
   \kern-.025em\copy0\kern-\wd0
   \kern.05em\copy0\kern-\wd0
   \kern-.025em\raise.0433em\box0 }
\def\sqr#1#2{{\vcenter{\vbox{\hrule height.#2pt
     \hbox{\vrule width.#2pt height#1pt \kern#1pt
   \vrule width.#2pt}\hrule height.#2pt}}}}

\def\e{\epsilon}

\def\e{\epsilon}
\def\s{\sigma}
\def\d{\delta}
\def\l{\lambda}

\def\g{\gamma}
\def\G{\Gamma}
\def\a{\alpha}
\def\b{\beta}

\def\cal{\mathcal}

\newenvironment{myenumerate}{%
\begin{list}{\labelenumi}
	{%
	\setlength{\itemsep}{0.4em}%
	\setlength{\topsep}{0.5em}%
	\setlength\leftmargin{2.6em}%
	\setlength\labelwidth{2.15em}%
	\setlength{\labelsep}{0.45em}%
	\usecounter{enumi}%
	}%
	}%
{\end{list}
}

\renewenvironment{enumerate}{
\renewcommand{\theenumi}{\arabic{enumi}}%
\renewcommand{\labelenumi}{{\rm(\theenumi)}}%
\begin{myenumerate}}%
{\end{myenumerate}}

\newenvironment{Aenumerate}{%
\renewcommand{\theenumi}{\Alph{enumi}}
\renewcommand{\labelenumi}{{\rm(\theenumi)}}
\begin{myenumerate}
	}%
{\end{myenumerate}}

\newenvironment{aenumerate}{%
\renewcommand{\theenumi}{\alph{enumi}}
\renewcommand{\labelenumi}{{\rm(\theenumi)}}
\begin{myenumerate}
	}%
{\end{myenumerate}}

\newenvironment{myitemize}{%
\begin{list}{$\bullet$}%
 	{%
	\setlength{\itemsep}{0.4em}%
	\setlength{\topsep}{0.5em}%
	\setlength\leftmargin{2.6em}%
	\setlength\labelwidth{2.15em}%
	\setlength{\labelsep}{0.45em}%
	}%
	}%
{\end{list}}

\renewenvironment{itemize}{
\begin{myitemize}}%
{\end{myitemize}}

\DeclareMathOperator*{\argmin}{arg\, min}
\def\dd{\mathrm{d}}
\def\cG{\mathcal{G}}
\def\gs{\sigma}
\def\cT{\mathcal{T}}
\def\cF{\mathcal{F}}

\def\cL{\mathcal{L}}

\title[Scaling and multiscaling in financial series]{Scaling and multiscaling 
in financial  series:\\ a simple model}

\author{Alessandro Andreoli}
\address{Dipartimento di Management,
	Universit\`a Politecnica delle Marche, 
	Piazzale Martelli 8,
	60121 Ancona
	Italy.}
\email{alessandro.andreoli@univpm.it}

\author{Francesco Caravenna}
\address{Dipartimento di Matematica e Applicazioni, 
Universit\`a degli Studi di Milano-Bicocca,
via Cozzi 53,
I-20125 Milano, Italy}

\email{francesco.caravenna@unimib.it}

\author{Paolo Dai Pra}
\address{Dipartimento di Matematica Pura ed Applicata,
	Universit\`a degli Studi di Padova,
	via Trieste 63,
	I-35121 Padova,
	Italy}
\email{daipra@math.unipd.it}

\author{Gustavo Posta}
\address{Dipartimento di Matematica,
	Politecnico di Milano,
	Piazzale Leonardo da Vinci 32,
	I-20133 Milano, Italy}
\email{gustavo.posta@polimi.it}



\thanks{We gratefully acknowledge the support of
the University of Padova under grant CPDA082105/08.}

\date{\today}

\begin{document}

\MakeRobustCommand\subref

\begin{abstract}
We propose a simple stochastic volatility model
which is analytically tractable, very easy to simulate and
which captures some relevant stylized facts of financial assets,
including \emph{scaling properties}.
In particular, the model displays a crossover in the log-return distribution 
from power-law tails (small time) to a Gaussian behavior (large time),
slow decay in the volatility autocorrelation and multiscaling of moments.
Despite its few parameters, the model is able to fit
several key features of the time series of
financial indexes, such as the \emph{Dow Jones Industrial Average},
with a remarkable accuracy.

\end{abstract}

\maketitle

\section{Introduction} 

\subsection{Modeling financial assets}

\label{sec:intro}

Recent developments in stochastic modelling of time series have been strongly 
influenced by the analysis of financial assets, such as exchange rates, stocks,
and market indexes.
The basic model, that has given 
rise to the celebrated Black \& Scholes  formula \cite{Hu,cf:KarShr}, assumes that the 
logarithm $X_t$ of the price of the asset, after subtracting the trend,
evolves through the simple equation
\be{1}
	\dd X_t = \s \, \dd B_t,
\ee
where $\s$ (the {\em volatility})
is a constant and $(B_t)_{t \geq 0}$ is a standard Brownian motion.
It has been well-know for a long time that, despite its success, this model 
is not consistent with a number of {\em stylized facts} that are empirically detected in many real time series, e.g.:
\begin{itemize}
\item
the volatility is not constant and may exhibit high peaks, 
that may be interpreted as \emph{shocks} in the market;
\item
the empirical distribution of the increments $X_{t+h} - X_t$ of the logarithm
of the price --- called {\em log-returns} --- is non Gaussian, 
displaying \emph{power-law tails}
(see Figure~\ref{fig:distribution}{\sc\subref{fig:dji_tails}} below),
especially for small values of the time span $h$,
while a Gaussian shape is approximately recovered for large values of $h$;
\item
log-returns corresponding to disjoint time-interval are uncorrelated, but not independent:
in fact, the correlation between the absolute values $|X_{t+h} - X_t|$ 
and $|X_{s+h} - X_s|$ --- called \emph{volatility autocorrelation} ---
is positive ({\em clustering of volatility})
and has a slow decay in $|t-s|$ ({\em long memory}),
at least up to moderate values for $|t-s|$
(cf. Figure~\ref{fig:comparisons}{\sc\subref{fig:compdji_corr400log}-\subref{fig:compdji_corr400loglog}} 
below).
\end{itemize}
In order to account for these facts,
a very popular choice in the literature of mathematical finance and financial economics
has been to upgrade the basic model \eqref{1},
allowing $\s = \s_t$ to vary with $t$ and to be itself a stochastic process.
This produces a wide class of processes, known as \emph{stochastic volatility models},
determined by the process $(\sigma_t)_{t\ge 0}$,
which are able to capture (at least some of) the above-mentioned stylized facts,
cf.~\cite{Ba:Sh,cf:SheAnd} and references therein.



\begin{figure}
\centering
\subfloat[][\emph{Diffusive scaling of log-returns}.]
{\includegraphics[width=.43\columnwidth]{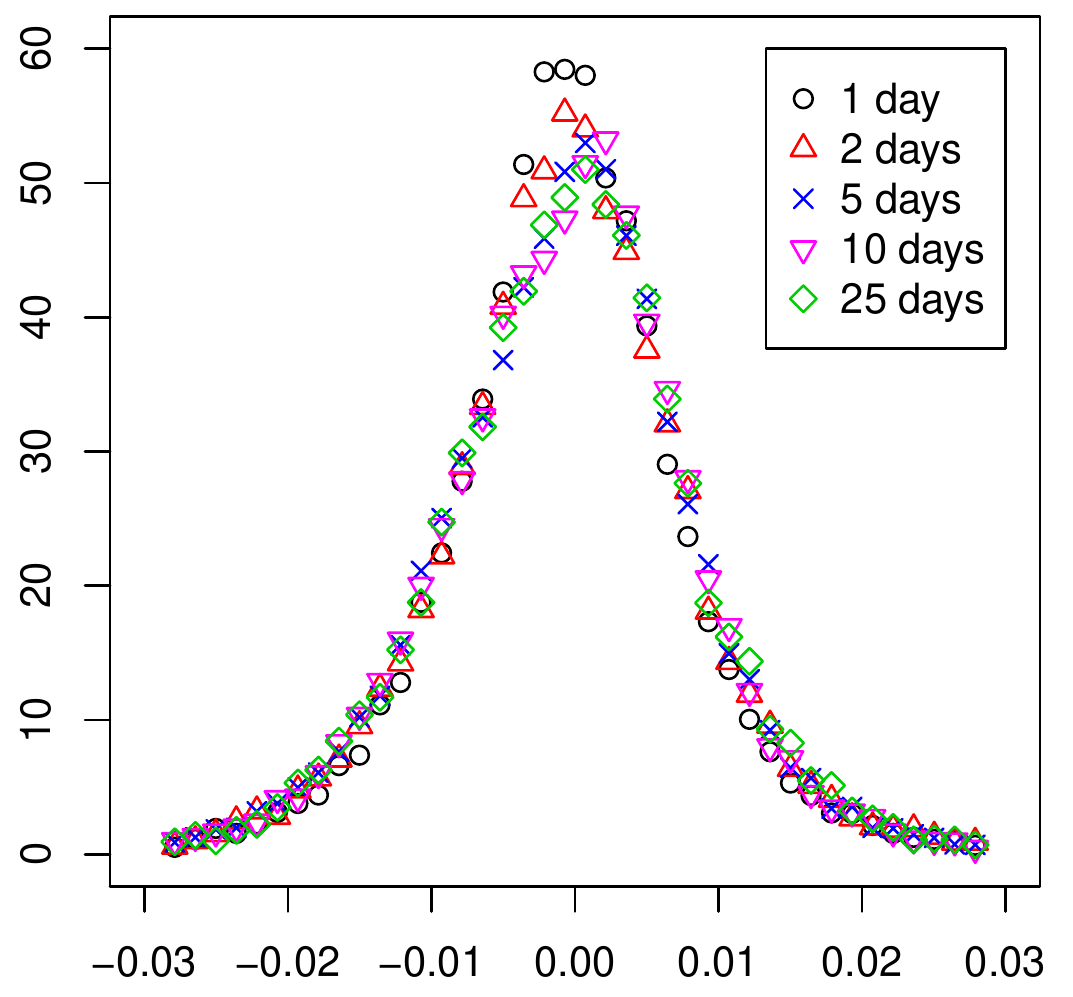}\label{fig:diffusive}}
\qquad \
\subfloat[][\emph{Multiscaling of moments}.]
{\includegraphics[width=.43\columnwidth]{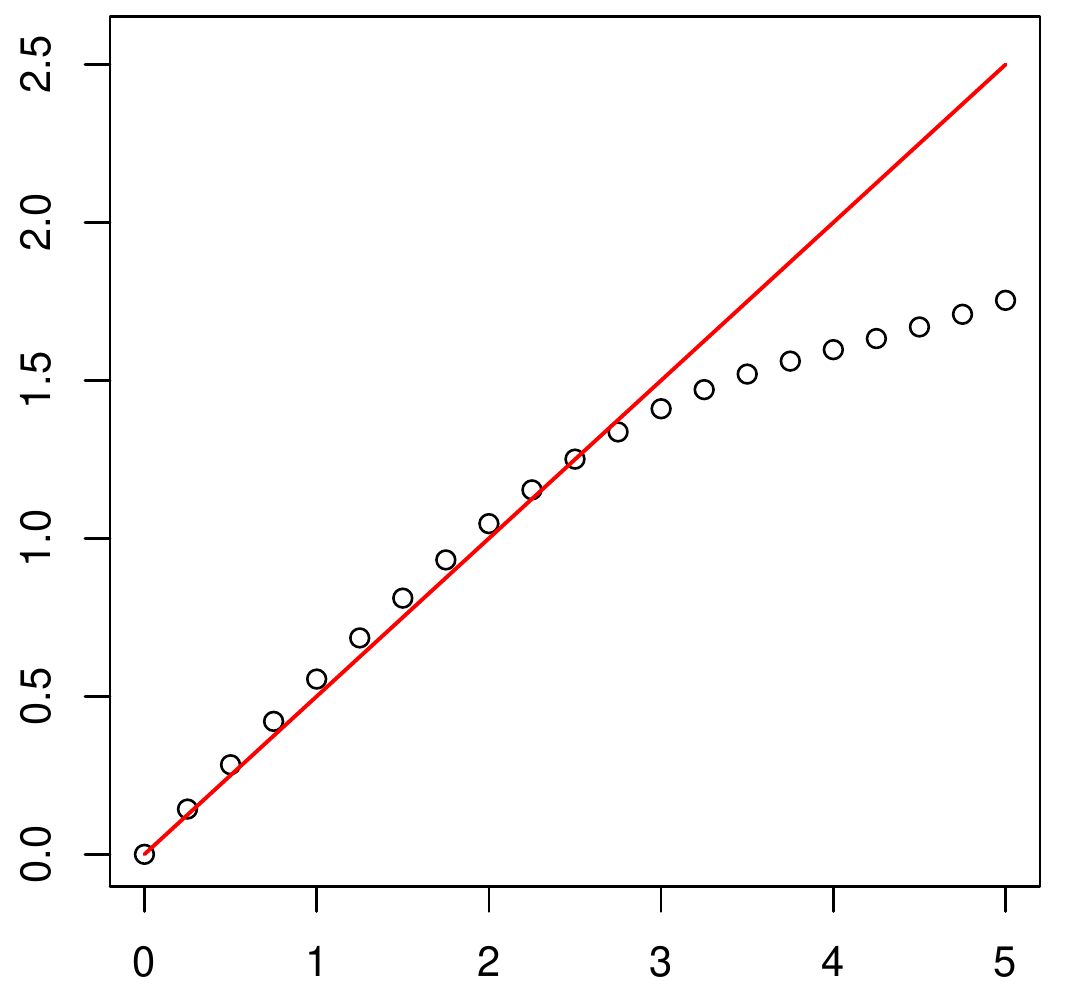}\label{fig:multiscaling1}}
\caption{\emph{Scaling properties of the DJIA time series (opening
prices 1935-2009)}.\\
{\sc\subref{fig:diffusive}}
The empirical densities of the log-returns over 1, 2, 5, 10, 25 days
show a remarkable overlap under diffusive scaling.\\
{\sc\subref{fig:multiscaling1}} The scaling exponent $A(q)$ as a function of $q$,
defined by the relation $m_q(h) \approx h^{A(q)}$ (cf. \eqref{eq:mq0}), bends down
from the Gaussian behavior $q/2$
(red line) for $q \ge \overline q \simeq 3$.
The quantity $A(q)$ is evaluated
empirically through a linear interpolation
of $(\log m_q(h))$ versus $(\log h)$ for $h \in \{1, \ldots, 5\}$
(cf. section~\ref{sec:numerics} for more details).} \label{fig:scaling}
\end{figure}

\smallskip

More recently, other stylized facts have been pointed out
concerning the \emph{scaling properties} 
of the empirical distribution of the log-returns. Given a daily
time series $(s_i)_{1 \leq i \leq T}$ over a period of $T \gg 1$ days,
denote by $p_h$ the \emph{empirical distribution} of the
(detrended) log-returns corresponding to an interval of $h$ days:
\be{eq1}
	p_h(\cdot) := \frac{1}{T-h} \sum_{i=1}^{T-h} \d_{x_{i+h} - x_i}(\cdot) \,,
	\qquad\ x_i := \log(s_i) - \overline{d}_i \,,
\ee
where $\overline{d}_i$ is the local rate of linear
growth of $\log (s_i)$ 
(see section~\ref{sec:numerics} for details)
and $\d_{x}(\cdot)$ denotes the Dirac measure
at $x \in \R$. The statistical analysis of various financial series, 
such as the \emph{Dow Jones Industrial Average} (DJIA) or the
\emph{Nikkei 225}, shows that, 
for small values of $h$, $p_h$ obeys approximately 
a \emph{diffusive scaling relation} (cf. Figure~\ref{fig:scaling}{\sc\subref{fig:diffusive}}):
\be{eq2}
p_h (\dd r) \simeq \frac{1}{\sqrt{h}} \, g\bigg( \frac{r}{\sqrt{h}} \bigg) \dd r,
\ee
where $g$ is a probability density with power-law tails.
Considering the $q$-th empirical moment $m_q(h)$, defined by
\begin{equation}\label{eq:mq0}
	m_q(h) \,:=\, \frac{1}{T-h} \sum_{i=1}^{T-h} |x_{i+h} - x_i|^q \,=\,
	\int |r|^q \, p_h(\dd r) \,,
\end{equation}
from relation \eqref{eq2} it is natural to guess that
$m_q(h)$ should scale as $h^{q/2}$. This is indeed what one
observes for moments of small order $q \leq \bar{q}$ (with $\bar{q} \simeq 3$ for the DJIA).
However, for moments of higher order $q > \bar{q}$, the different scaling relation
$h^{A(q)}$, with $A(q) < q/2$, takes place,
cf. Figure~\ref{fig:scaling}{\sc\subref{fig:multiscaling1}}.
This is the so-called \emph{multiscaling of moments},
cf. \cite{cf:Vas,cf:Gha,cf:Gal,DiM}.

An interesting class of models that are able to reproduce the multiscaling of moments
--- as well as many other features, notably the persistence of volatility on different
time scales --- are the so-called \emph{multifractal models}, like the
MMAR (Multifractal Model of Asset Returns, cf. \cite{cf:CalFisMan})
and the MSM (Markov-Switching Multifractal, cf. \cite{cf:CalFis}).
These models describe the evolution of the detrended log-price $(X_t)_{t\ge 0}$ 
as a \emph{random time-change of Brownian motion}:
$X_t = W_{I(t)}$, where the time-change $(I(t))_{t\ge 0}$ is a continuous and 
increasing process, sometimes called \emph{trading time},
which displays \emph{multifractal features} and is usually 
taken to be independent of the Brownian motion $(W_t)_{t\ge 0}$
(cf. \cite{cf:CalFisBook} for more details).

Modeling financial series through a random time-change of Brownian motion
is a classical topic, dating back to Clark \cite{Cl}, and reflects the natural idea that
external information influences the speed at which exchanges take place
in a market. It should be stressed that, under the mild regularity
assumption that the time-change $(I(t))_{t\ge 0}$ 
has absolutely continuous paths a.s., any random 
time-change of Brownian motion $X_t = W_{I(t)}$ can be written as
a stochastic volatility model
$\dd X_t = \sigma_t \, \dd B_t$, and viceversa.\footnote{More precisely, 
``independent random time changes of Brownian motion with
absolutely continuous time-change''
--- that is, processes $(X_t)_{t\ge 0}$ such that
$X_t - X_0 = W_{I(t)}$, where $(W_t)_{t\ge 0}$ is a Brownian motion
and $(I(t))_{t\ge 0}$ is an independent process with increasing and absolutely
continuous paths a.s. --- and ``stochastic volatility models
with independent volatility'' --- that is, processes $(X_t)_{t\ge 0}$ such that
$\dd X_t = \sigma_t \, \dd B_t$, where $(B_t)_{t\ge 0}$ is a Brownian motion 
and $(\sigma_t)_{t\ge 0}$ is an independent process with paths in $L^2_{\mathrm{loc}}(\R)$ a.s. ---
are the same class of processes,
cf. \cite{Ba:Sh,cf:SheAnd}. The link between the two representations
$\dd X_t = \sigma_t \, \dd B_t$ and $X_t - X_0 = W_{I(t)}$ is given by $\sigma_t = \sqrt{I'(t)}$ and
$B_t = \int_0^t (I'(s))^{-1/2} \dd W_{I(s)} = \int_0^{I_t} (I'(I^{-1}(v)))^{-1/2} \dd W_u$.} 
However, a key feature of multifractal models is precisely that
their trading time $(I(t))_{t\ge 0}$ has \emph{non absolutely continuous paths} a.s.,
hence they cannot be expressed as stochastic volatility models.
This makes their analysis harder, as the standard tools
available for Ito diffusions cannot be applied.

%

\smallskip

The purpose of this paper is to define a
\emph{simple} stochastic volatility model --- or, equivalently,
an independent random time change of Brownian motion,
where the time-change process has absolutely continuous paths --- which agrees with
all the above mentioned stylized facts, displaying in particular
a crossover in the log-return distribution from a power-law to a Gaussian behavior,
slow decay in the volatility autocorrelation, diffusive scaling and multiscaling of moments.
In its most basic version, the model contains only three real parameters
and is defined as a simple, deterministic function of a Brownian motion and an independent
Poisson process. This makes the process analytically tractable and very easy to simulate.
Despite its few degrees of freedom, the model is able to fit
remarkably well several key features of
the time series of the main financial indexes, such as
DJIA, S\&P~500, FTSE~100, Nikkei~225. In this paper
we present a detailed numerical analysis on the DJIA.

Let us mention that there are subtler stylized facts that are not properly
accounted by our model, such as 
the multi-scale intermittency of the volatility profile,
the possible skewness of the log-return distribution and
the so-called \emph{leverage effect}
(negative correlation between log-returns and future volatilities),
cf.~\cite{cf:Cont}.
As we discuss in section~\ref{sec:discussion},
such features --- that are relevant in the analysis of particular assets ---
can be incorporated in our model in a natural way.
Generalizations in this sense are currently under investigation,
as are the performances of our model in financial problems, 
like the pricing of options (cf. A. Andreloli's Ph.D. Thesis~\cite{cf:Andreoli}).
In this article we stick for simplicity to the most basic formulation.

\smallskip

Finally, although we work in the framework of stochastic volatility models, we point out
that an important alternative class of models in discrete time,
widely used in the econometric literature, is given by
autoregressive processes such as ARCH, GARCH, FIGARCH and their generalizations,
cf. \cite{En,Bo, Ba, BoMi}. More recently, continuous-time versions
have been studied as well, cf.~\cite{Kl:Li:Ma,Kl:Li:Ma:1}.
With no aim for completeness, let us mention that 
GARCH and FIGARCH do not display multiscaling of moments,
cf. \cite[\S8.1.4]{cf:CalFisBook}. 
We have also tested the model recently proposed 
in \cite{Bo:Kr:Pi:Ta}, which is extremely accurate to fit the statistics of the empirical volatility,
and it does exhibit multiscaling of moments. The price to pay is, however, that the
model requires the calibration of more than 30 parameters.

We conclude noting that long memory effects in autoregressive models are
obtained through a suitable dependence on the past in the equation for the volatility, 
while large price variations are usually controlled by specific features of the driving noise.
In our model, we propose a \emph{single} mechanism, 
modeling the reaction of the market to shocks, which is the source of all the
mentioned stylized facts.


\smallskip

\subsection{Content of the paper}
The paper is organized as follows.
\begin{itemize}
\item In section~\ref{sec:model} we give the definition of our model,
we state its main properties and we discuss its ability to fit the DJIA time series
in the period 1935-2009.

\item In section~\ref{sec:discussion} we discuss some key features
and limitations of our model, point out possible generalizations, and compare it
with other models.

\item Sections~\ref{sec:multi}, \ref{sec:corr} and~\ref{sec:basic}
contain the proofs of the main results, plus some additional material.

\item In section~\ref{sec:numerics} we discuss more in detail
the numerical comparison between our model and the
DJIA time series.

\item Finally, appendix~\ref{sec:appetails} contains the proof of some
technical results, while
appendix~\ref{sec:BalSte} is devoted to a brief discussion of the
model introduced by F.~Baldovin and A.~Stella in~\cite{BV1,BV2}, which has
partially inspired the construction of our model.
\end{itemize}

\subsection{Notation}

Throughout the paper, the indexes $s,t,u,x,\lambda$ run over real
numbers while $i,k,m,n$ run over integers, so that
$t \ge 0$ means $t \in [0,\infty)$ while $n \ge 0$ means
$n \in \{0,1,2,\ldots\}$. The symbol ``$\sim$'' denotes
asymptotic equivalence for positive sequences ($a_n \sim b_n$ if and only if
$a_n/b_n \to 1$ as $n\to\infty$) and also equality in law for random variables,
like $W_1 \sim \mathcal{N}(0,1)$.
Given two real functions $f(x)$ and $g(x)$, we write
$f = O(g)$ as $x \to x_0$ if there exists $M > 0$ such that
$|f(x)| \le M |g(x)|$ for $x$ in a neighborhood of $x_0$,
while we write $f = o(g)$ if $f(x)/g(x) \to 0$ as $x \to x_0$; in particular,
$O(1)$ (resp. $o(1)$) is a bounded (resp. a vanishing) quantity.
The standard exponential and Poisson laws
are denoted by $Exp(\lambda)$ and $Po(\lambda)$, for $\lambda > 0$:
$X \sim Exp(\lambda)$ means
that $P(X\le x) = (1-e^{-\lambda x})\ind_{[0,\infty)}(x)$ for all $x \in \R$
while $Y \sim Po(\lambda)$ means that $P(Y = n) = e^{-\lambda} \lambda^n / n!$
for all $n \in \{0,1,2,\ldots\}$.
We sometimes write $(const.)$ to denote a positive constant, whose value
may change from place to place.


\medskip
\section{The model and the main results}

\label{sec:model}

We introduce our model as an {\em independent random time change} of a Brownian motion, 
in the spirit  e.g. of \cite{Cl} and \cite{An:Ge}. 
An alternative and equivalent definition, as a stochastic volatility model, 
is illustrated in section~\ref{subsec:basic}.

\subsection{Definition of the model}

In its basic version, our model contains only three real parameters:
\begin{itemize}
\item $\l \in (0,+\infty)$ is the inverse of the average waiting time between ``shocks'' in the market;

\item $D \in (0,1/2]$ determines the sub-linear time change $t \mapsto t^{2D}$, 
which expresses the ``trading time'' after shocks; 

\item $\s \in (0,+\infty)$ is proportional to the average volatility.
\end{itemize}
In order to have more flexibility, we actually let $\s$ be a \emph{random} parameter, 
i.e., a positive random variable, whose distribution $\nu$ becomes the relevant parameter:
\begin{itemize}
\item $\nu$ is a probability on $(0,\infty)$, \emph{connected} to the volatility distribution.
\end{itemize}

\begin{remark}
When the model is calibrated to the main financial indexes
(DJIA, S\&P~500, FTSE~100, Nikkei~225), the best fitting turns out to be obtained for a 
\emph{nearly constant} $\s$. In any case, we stess that
the main properties of the model are only marginally dependent on the law $\nu$ of $\sigma$:
in particular, the first two moments of $\nu$, i.e. $E(\sigma)$ and $E(\sigma^2)$,
are enough to determine the features of our model that are
relevant for real-world times series, cf. Remark~\ref{rem:robustness} below.
Therefore, roughly speaking, we could say
that in the general case of random $\sigma$ our model has four
``effective'' real parameters.
\end{remark}

Beyond the parameters $\lambda, D, \nu$,
we need the following three sources of alea:
\begin{itemize}
\item a standard Brownian motion $W = (W_t)_{t\ge 0}$;
\item a Poisson point process $\cal T = (\tau_n)_{n \in \Z}$ on $\R$ 
with intensity $\lambda$;
\item a sequence $\Sigma = (\s_n)_{n \ge 0}$
of independent and identically distributed positive random variables with law $\nu$ (so that $\gs_n \sim \nu$
for all $n$);
and for conciseness we denote by $\sigma$ a variable with the same law $\nu$.
\end{itemize}
We assume that $W, \cal T, \Sigma$ are defined on some probability
space $(\Omega, \cal F, P)$ and that they are independent.
By convention, we label the points of $\cT$ so that $\tau_0 < 0 < \tau_1$.
We will actually need only the points $(\tau_n)_{n \ge 0}$,
and we recall that the random variables
$(-\tau_0)$, $\tau_1$, $(\tau_{n+1} - \tau_n)_{n\ge 1}$
are independent and identically distributed with marginal laws $Exp(\lambda)$.
In particular, $1/\lambda$ is the mean distance between the points in $\cT$,
except for $\tau_0$ and $\tau_1$, whose average distance is $2/\lambda$.
Although some of our results would hold for 
more general distributions of ${\cal{T}}$, we stick
for simplicity to the (rather natural) choice of a Poisson process.

\begin{figure}
\centering
\includegraphics[width=.65\columnwidth]{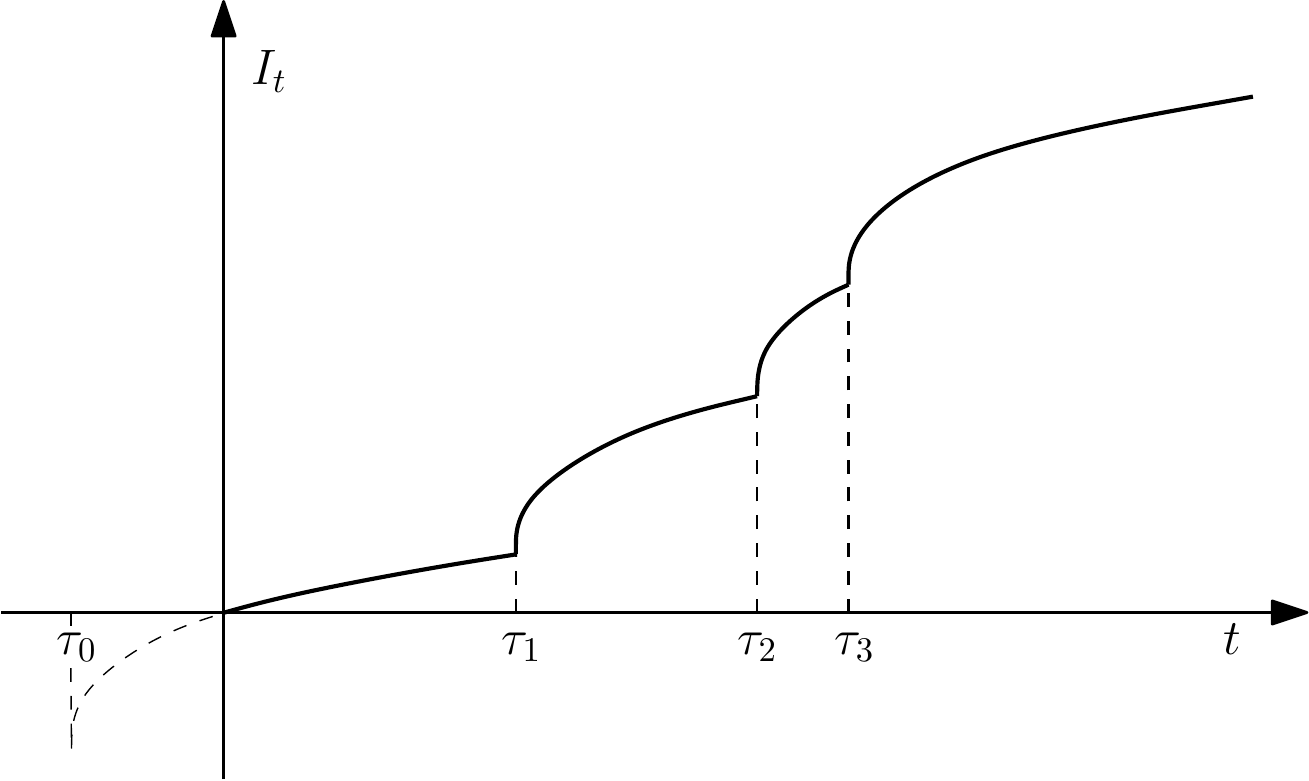}
\caption{A sample trajectory of the process $(I_t)_{t\ge 0}$\label{fig:It}}
\end{figure}

For $t \ge 0$, 
we define
\begin{equation} \label{eq:it}
i(t) \,:=\, \sup\{n \ge 0 :\, \tau_n \leq t\} \,=\, \#\{ \cT \cap (0,t] \} \,,
\end{equation}
so that $\tau_{i(t)}$ is the location of the last point
in $\cal T$ before~$t$. Plainly, $i(t) \sim Po(\lambda t)$.
Then we introduce the basic process
$I = (I_t)_{t\ge 0}$ defined by
\be{eq10}
I_t \,=\, I(t) \,:=\, \s_{i(t)}^2 \left(t - \tau_{i(t)} \right)^{2D} \,+\,
\sum_{k=1}^{i(t)} \s_{k-1}^2 \left(\tau_{k} - \tau_{k-1} \right)^{2D} \,-\,
\s^2_0 \left(-\tau_0\right)^{2D},
\ee
with the agreement that the sum in the right hand side is
zero if $i(t) = 0$. 
More explicitly, $(I_t)_{t\ge 0}$ is a continuous process with $I_0 = 0$
and $I_{\tau_n + h} - I_{\tau_n} = (\sigma_n^2)\, h^{2D}$ for 
$0 \le h \le (\tau_{n+1}-\tau_n)$.
We note that the derivative $(\frac{\dd}{\dd t}I_t)_{t\ge 0}$ is
a \emph{stationary regenerative process}, cf.~\cite{cf:Asm}.
See Figure~\ref{fig:It} for a sample trajectory of $(I_t)_{t\ge 0}$
when $D < \frac{1}{2}$.

We then define our
model $X = (X_t)_{t\ge 0}$ by setting
\be{eq11}
X_t := W_{I_t} \,.
\ee
In words: our model is an random time change of the Brownian motion
$(W_t)_{t\ge 0}$ through the time-change process $(I_t)_{t\ge 0}$.
Note that $I$ is a strictly increasing process with absolutely continuous paths,
and it is independent of $W$. 

When $D = \frac{1}{2}$ and $\sigma$ is constant, we have $I_t = \sigma^2 t$ and
the model reduces to Black \& Scholes with volatility $\s$.
On the other hand, when $D<\frac12$, the paths of $I$ are singular (non differentiable)
at the points in ${\cal{T}}$, cf. Figure~\ref{fig:It}.
This suggests a possible financial interpretation
of the instants in ${\cal{T}}$ as the epochs at which
{\em big shocks} arrive in the market, making the volatility jump to infinity. 
This will be more apparent in next subsection, where we 
give a stochastic volatility formulation of the model.
We point out that the singularity is produced by the sub-linear time change $t \mapsto t^{2D}$, 
that was first suggested by F. Baldovin and A. Stella in~\cite{BV1,BV2}
(their model is described in appendix~\ref{sec:BalSte}).

\subsection{Basic properties}
\label{subsec:basic}

Let us state some basic properties of our model, that will be proved
in section~\ref{sec:basic}.

\begin{Aenumerate}
\item \label{prop:stationarity}
{\em The process $X$ has stationary increments}. 

\item \label{prop:momentsXsigma}
{\em The following relation between moments of $X_t$ and $\sigma$ holds}:
for any $q > 0$
\begin{equation} \label{eq:qXsigma}
	E(|X_t|^q) < \infty \text{ for some (hence any) } t > 0
	\quad\Longleftrightarrow\quad E(\sigma^q) < \infty \,.
\end{equation}

\item \label{prop:randomvol}
{\em The process $X$ 
can be represented as a stochastic volatility model}:
\begin{equation} \label{stochvol}
	\dd X_t = v_t\, \dd B_t \,,
\end{equation}
where $(B_t)_{t\ge 0}$ is a standard Brownian motion
and $(v_t)_{t\ge 0}$ is an independent process, defined by
(denoting $I'(s) := \frac{\dd}{\dd s}I(s)$)
\be{volatility}
\begin{split}
	B_t & := \int_0^{t} 
	\frac{1}{\sqrt{I'(s)}} \, \dd W_{I(s)} = \int_0^{I_t} 
	\frac{1}{\sqrt{I'(I^{-1}(u))}} \, \dd W_u \,, \\
	v_t & := \sqrt{ I'(t) } = \sqrt{2D} \, \s_{i(t)}
	\left(t - \tau_{i(t)} \right)^{D- \frac{1}{2}} \,.
\end{split}
\ee
Note that, whenever $D < \frac{1}{2}$, the volatility $v_t$ has singularities at the random times 
$\tau_n$.

\item \label{prop:martingale}
{\em The process $X$ is a zero-mean, square-integrable martingale}
(provided $E(\sigma^2) < \infty$).
\end{Aenumerate}

\begin{remark}
If we look at the process $X$ for a \emph{fixed} realization
of the variables $\cT$ and $\Sigma$, averaging only
on $W$ --- that is, if we work under
the conditional probability $P(\,\cdot\,|\cT, \Sigma)$ ---
the increments of $X$ are no longer stationary, but the properties 
\eqref{prop:randomvol} and \eqref{prop:martingale}
continue to hold (of course, condition $E(\sigma^2) < \infty$
in \eqref{prop:martingale} is not required under $P(\,\cdot\,|\cT, \Sigma)$).
\end{remark}

\begin{remark}\label{rem:realtails}
It follows from \eqref{eq:qXsigma} that
if $\s$ is chosen as a deterministic constant, then $X_t$ admits 
moments of all order (actually, even exponential moments, cf. Proposition~\ref{prop:etails}
in section~\ref{sec:basic}). 
This seems to indicate that to \emph{see} power-law tails in the distribution
of $(X_{t+h} - X_t)$ --- one of the basic stylized facts mentioned in the introduction ---
requires to take $\sigma$ with power-law tails.
\emph{This, however, is not true}, and is one on the surprising features 
of the simple model we propose: for typical choices of the parameters of our model,
the distribution of $(X_{t+h} - X_t)$
displays a power-law tail behavior up to several standard deviations from the mean,
irrespective of the law of $\sigma$.
Thus, the eventually light tails are ``invisible'' for all practical purposes
and real heavy-tailed distributions appear to be unnecessary to fit data. 
We discuss this issue below, cf. Remark~\ref{rem:tails}, after having stated some results;
see also subsection~\ref{subsec:fitting} and Figure~\ref{fig:distribution}{\sc\subref{fig:dji_tails}}
for a graphical comparisons with the DJIA time series.
\end{remark}

\smallskip

Another important property of the process $X$ is that
its increments are \emph{mixing},
as we show in section~\ref{sec:basic}. This
entails in particular that for
every $\delta > 0$, $k \in \N$ and for every choice of
the intervals $(a_1, b_1)$, \ldots, $(a_k, b_k) \subseteq (0,\infty)$
and of the measurable function $F: \R^k \to \R$, we have almost surely
\begin{equation} \label{eq:erg}
\begin{split}
	& \lim_{N\to\infty} \, \frac 1N \sum_{n=0}^{N-1}
	F(X_{n\delta + b_1} - X_{n\delta + a_1},\, \ldots,\,
	X_{n\delta + b_k} - X_{n\delta + a_k}) \\
	& \qquad \qquad \qquad \qquad \qquad 
	\qquad \qquad = E\big[ F(X_{b_1} - X_{a_1},\, \ldots,\, X_{b_k} - X_{a_k}) \big] \,,
\end{split}
\end{equation}
provided the expectation appearing in the right hand
side is well defined. In words: the empirical average
of a function of the increments of the process
over a long time period is close to its expected value.

Thanks to this property, our main results concerning the
distribution of the increments of the process $X$,
that we state in the next subsection,
are of direct relevance for the comparison of our model
with real data series. Some care is needed, however,
because the accessible time length $N$ in \eqref{eq:erg} may not be large
enough to ensure that the empirical averages
are close to their limit. We elaborate more on
this issue in section~\ref{sec:numerics}, where we
compare our model with the DJIA data from a numerical viewpoint.

\subsection{The main results}

We now state our main results for our model $X$,
that correspond to the basic stylized facts
mentioned in the introduction: diffusive scaling and crossover of the 
log-return distribution (Theorem~\ref{th:scaling}); 
multiscaling of moments (Theorem~\ref{th:multi} and Corollary~\ref{A(q)}); 
clustering of volatility (Theorem~\ref{th:cov} and Corollary~\ref{cor:volauto}).

\smallskip

Our first result, proved in section~\ref{sec:multi},
shows that the increments $(X_{t+h} - X_t)$ have an approximate diffusive scaling
both when $h\downarrow 0$, with a heavy-tailed limit distribution
(in agreement with \eqref{eq2}), and when $h \uparrow \infty$,
with a normal limit distribution. This is a precise mathematical
formulation of a crossover phenomenon in the log-return distribution,
from approximately heavy-tailed (for small time) to approximately Gaussian (for large time).

\begin{theorem}[Diffusive scaling] \label{th:scaling}
The following convergences in distribution hold
for any choice of the parameters $D, \lambda$ and of the law $\nu$ of $\sigma$.
\begin{itemize}
\item Small-time diffusive scaling:
\begin{align} \label{eq:convdist}
	\frac{(X_{t+h} - X_t)}{\sqrt h} \ \xrightarrow[\ h \downarrow 0\ ]{d} \
	f(x) \, \dd x \ := \ \text{law of } \ \big(\sqrt{2D} \,
	\lambda^{\frac{1}{2} - D} \big) \, \s \, S^{D-\frac{1}{2}} \, W_1 \,,
\end{align}
where $\sigma \sim \nu$, 
$S \sim Exp(1)$ and $W_1 \sim \mathcal{N}(0,1)$
are independent random variables.
The density $f$ is thus a mixture of centered Gaussian densities
and, when $D < \frac{1}{2}$, has \emph{power-law tails}:
more precisely, if $E(\sigma^q) < \infty$ for all $q > 0$,
\begin{equation} \label{eq:fqstar}
	\int |x|^q f(x) \, \dd x \, < +\infty \ \iff \ q < q^* := \frac{1}{(\frac{1}{2} - D)}.
\end{equation}


\item Large-time diffusive scaling: if $E(\sigma^2) < \infty$
\begin{align} \label{eq:convdist2}
	\frac{(X_{t+h} - X_t)}{\sqrt h} \ \xrightarrow[\ h \uparrow \infty\ ]{d} \
	\frac{e^{-x^2/(2c^2)}}{\sqrt{2\pi} c} \, \dd x \, = \, \mathcal{N}(0,c^2) \,,
	\qquad
	c^2 = \lambda^{1-2D} \, E(\sigma^2) \, \Gamma(2D+1) \,,
\end{align}
where $\Gamma(\alpha) := \int_0^\infty x^{\alpha-1} e^{-x} \dd x$
denotes Euler's Gamma function.
\end{itemize}
\end{theorem}

\begin{remark}
\label{rem:tails}
We have already observed that, when $\sigma$ has finite moments of all orders,
for $h>0$ the increment $(X_{t+h} - X_t)$ has finite moments of all orders too,
cf. \eqref{eq:qXsigma}, so there are no heavy tails in the strict sense. 
However, for $h$ small, the heavy-tailed density $f(x)$ is
by \eqref{eq:convdist} an excellent approximation for
the true distribution of $\frac{1}{\sqrt h} (X_{t+h} - X_t)$ up to a certain distance from the mean,
which can be quite large.
For instance, when the parameters of our model are calibrated to the DJIA time series, 
these ``apparent power-law tails'' are clearly visible
for $h=1$ (daily log-returns) up to a distance of about \emph{six standard 
deviations from the mean}, cf. subsection~\ref{subsec:fitting} and
Figure~\ref{fig:distribution}{\sc\subref{fig:dji_tails}} below.

We also note that the moment condition \eqref{eq:fqstar} follows immediately from \eqref{eq:convdist}:
in fact, when $\sigma$ has finite moments of all orders,
\begin{equation} \label{eq:fqstar2}
	\int |x|^q f(x) \, \dd x \, < +\infty \ \iff \ E\big[S^{(D-1/2)q}\big] 
	=  \int_{0}^\infty  s^{(D-1/2)q} \, e^{- s} \, \dd s \, < +\infty\,,
\end{equation}
which clearly happens if and only if $q < q^* := (\frac{1}{2} - D)^{-1}$.
This also shows that the heavy tails of $f$ depend on the fact that 
the density of the random variable $S$, which represents (up to a constant)
the distance between points in $\cT$, is strictly 
positive around zero, and not on other details of the exponential distribution.
\end{remark}

The power-law tails of $f$ have striking consequences on the
scaling behavior of the moments of the increments of our model.
If we set for $q \in (0,\infty)$
\begin{equation} \label{eq:mq}
	m_q(h) := E(|X_{t+h} - X_t|^q) \,,
\end{equation}
the natural scaling $m_q(h) \approx h^{q/2}$ as $h\downarrow 0$, 
that one would naively guess from \eqref{eq:convdist}, breaks down
for $q > q^*$, when the faster scaling $m_q(h) \approx h^{Dq + 1}$ holds
instead, the reason being precisely the fact that the
$q$-moment of $f$ is infinite for $q\ge q^*$.
More precisely, we have the following result, that we prove
in section~\ref{sec:multi}. 

\begin{theorem}[Multiscaling of moments] \label{th:multi}
Let $q>0$, and assume $E\left(\s^q \right) < +\infty$.
The quantity $m_q(h)$ in \eqref{eq:mq} 
is finite and has the
following asymptotic behavior as $h \downarrow 0$:
\begin{equation*}
	m_q(h) \,\sim\, \begin{cases}
	C_q \, h^{\frac q2} & \text{if } q < q^*\\
	\rule{0pt}{1.2em}C_q \, h^{\frac q2} \log(\frac 1h) & \text{if } q = q^*\\
	\rule{0pt}{1.2em}C_q \, h^{Dq+1} & \text{if } q > q^*
	\end{cases}\,, \qquad
	\text{where } \ q^* := \frac{1}{(\frac 12 - D)} \,.
\end{equation*}
The constant $C_q \in (0,\infty)$ is given by
\begin{equation} \label{eq:Cq}
	C_q \,:=\, \begin{cases}
	E(|W_1|^q) \, E(\s^q) \, \lambda^{q/q^*}  \, (2D)^{q/2}
	\, \Gamma(1-q/q^*)  & \text{if } q < q^*\\
	\rule{0pt}{1.2em} E(|W_1|^q) \, 
	E(\s^{q}) \, \l \, (2D)^{q/2} & \text{if } q = q^*\\
	\rule{0pt}{1.2em}E(|W_1|^q) \, E(\s^q) \, \l \, 
	\big[\int_0^\infty ((1+x)^{2D} - x^{2D})^{\frac q2} \, \dd x 
	\, + \, \frac{1}{Dq+1}\big]  & \text{if } q > q^*
\end{cases}\,,
\end{equation}
where 
$\Gamma(\alpha) := \int_0^\infty x^{\alpha-1} e^{-x} \dd x$
denotes Euler's Gamma function.
\end{theorem}

\begin{corollary} \label{A(q)}
The following relation holds true:
\begin{equation} \label{eq:A(q)}
	A(q) := \lim_{h \downarrow 0} \frac{\log m_q(h)}{\log h} =
	\begin{cases}
	\displaystyle \frac{q}{2} & \text{if } q \le q^* \\
	\rule{0pt}{1.4em}
	Dq+1 & \text{if } q \ge q^*
	\end{cases} \,,
	\quad \ \
	\text{where } \ q^* := \frac{1}{(\frac 12 - D)} \,.
\end{equation}
\end{corollary}

\noindent
The explicit form \eqref{eq:Cq} of the multiplicative constant $C_q$
will be used in section~\ref{sec:numerics} for the estimation of
the parameters of our model on the DJIA time series.

\smallskip

Our last theoretical result, proved in section
\ref{sec:corr}, concerns the correlations of the absolute value of two increments,
usually called \emph{volatility autocorrelation}.
We start determining the behavior of the covariance.

\begin{theorem} \label{th:cov}
Assume that $E(\gs^2) < \infty$.
The following relation holds as $h\downarrow 0$, for all $s,t > 0$:
\begin{equation} \label{eq:cov}
	Cov(|X_{s+h} - X_s|,|X_{t+h}-X_{t}|)
	= \frac{4D}{\pi}
	\, \l^{1-2D} \, e^{-\lambda|t-s|} \, \big( \phi( \lambda |t-s|) \,
	h \,+\, o(h) \big) \,,
\end{equation}
where 
\begin{equation} \label{eq:phi}
	\phi(x) := Cov\big( \sigma \,S^{D-1/2} \,,
	\, \sigma \, \big( S + x \big)^{D-1/2} \big)
\end{equation}
and $S \sim Exp(1)$ is independent of $\sigma$.
\end{theorem}

\noindent
We recall that $\rho(Y,Z) := Cov(Y,Z)/\sqrt{Var(Y) Var(Z)}$
is the correlation coefficient of two random variables $Y,Z$.
As Theorem~\ref{th:multi} yields
\begin{equation*}
	\lim_{h\downarrow 0} \frac 1h \, Var(|X_{t+h}-X_t|) 
	\,=\, (2D) \, \lambda^{1-2D} \,
	Var(\sigma\, |W_1|\, S^{D-1/2}) \,,
\end{equation*}
where $S \sim Exp(1)$ is independent of $\sigma, W_1$,
we easily obtain the following result.

\begin{corollary}[Volatility autocorrelation] \label{cor:volauto}
Assume that $E(\gs^2) < \infty$.
The following relation holds as $h\downarrow 0$, for all $s,t > 0$:
\begin{equation} \label{eq:volauto}
\begin{split}
	& \lim_{h\downarrow 0} \, \rho(|X_{s+h} - X_s|,|X_{t+h}-X_{t}|) \\
	&  \qquad \qquad \,=\, \rho(t-s) 
	\,:=\, \frac{2}{\pi \, Var(\sigma\, |W_1|\, S^{D-1/2})} \, e^{-\lambda |t-s|}
	\, \phi( \lambda |t-s|) \,,
\end{split}
\end{equation}
where $\phi(\cdot)$ is defined in
\eqref{eq:phi} and 
$\sigma \sim \nu$, 
$S \sim Exp(1)$, $W_1 \sim \mathcal{N}(0,1)$
are independent random variables.
\end{corollary}

\smallskip

This shows that the volatility autocorrelation of our process
decays exponentially fast for time scales larger than the
mean distance $1/\lambda$ between the epochs $\tau_k$.
However, for shorter time scales
the relevant contribution is given by the
function $\phi(\cdot)$. By \eqref{eq:phi} we can write
\begin{equation} \label{eq:phi2}
	\phi(x) \;=\; Var(\s) \, E(S^{D-1/2}\, (S+x)^{D-1/2}) \,+\,
	E(\s)^2 \, Cov(S^{D-1/2}, (S+x)^{D-1/2}) \,,
\end{equation}
where $S \sim Exp(1)$. When $D < \frac{1}{2}$,  as $x\to\infty$ the two terms
in the right hand side decay as
\begin{equation} \label{eq:phi3}
	E(S^{D-1/2}\, (S+x)^{D-1/2}) \,\sim\, c_1 \, x^{D-1/2} \,,
	\quad
	Cov(S^{D-1/2}, (S+x)^{D-1/2}) \,\sim\, c_2 \, x^{D-3/2} \,.
\end{equation}
where $c_1, c_2$ are positive constants,
hence $\phi(x)$ has a power-law behavior as $x \to\infty$.
For $x = O(1)$, which is the relevant regime, the decay of $\phi(x)$
is, roughly speaking, slower than exponential
but faster than polynomial
(see Figures~\ref{fig:comparisons}{\sc\subref{fig:compdji_corr400log}}
and~\ref{fig:comparisons}{\sc\subref{fig:compdji_corr400loglog}}).

\subsection{Fitting the DJIA time series} \label{subsec:fitting}

\begin{figure}
\centering
\subfloat[][\emph{Multiscaling in the DJIA (1935-2009)}.]
{\includegraphics[width=.43\columnwidth]{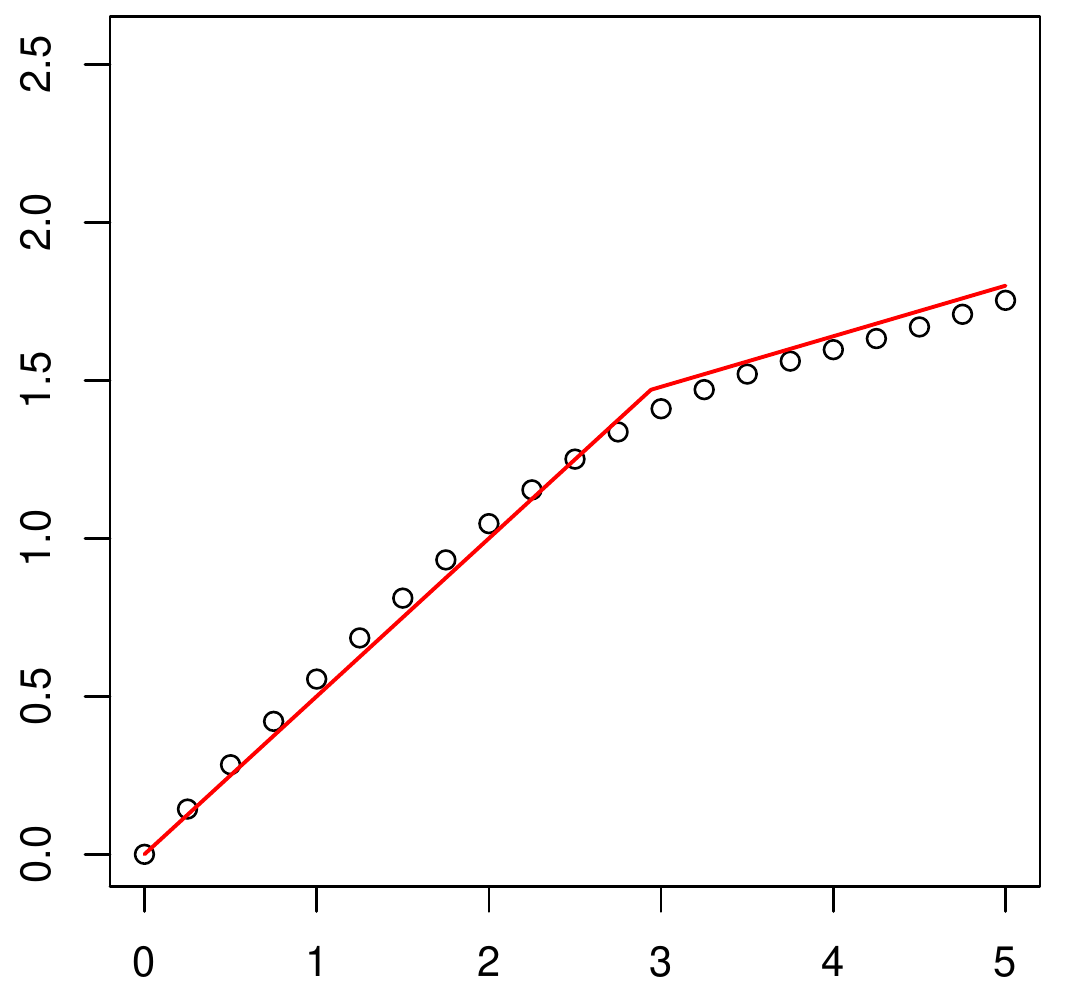}\label{fig:compmultiscaling2}}\\
\smallskip
\subfloat[][\emph{Volatility autocorrelation in the DJIA (1935-2009): log plot}.]
{\includegraphics[width=.43\columnwidth]{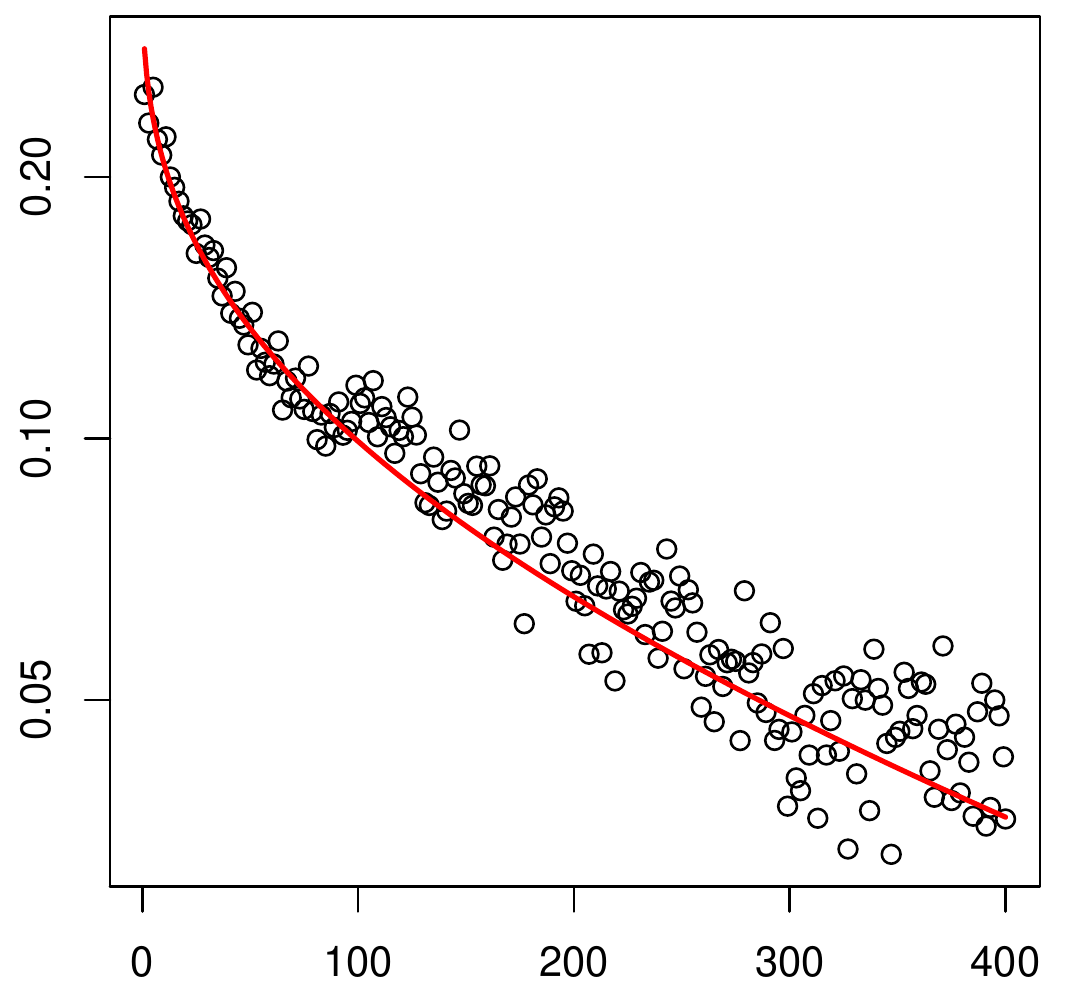}\label{fig:compdji_corr400log}}
\qquad \
\subfloat[][\emph{Volatility autocorrelation in the DJIA (1935-2009): log-log plot}.]
{\includegraphics[width=.43\columnwidth]{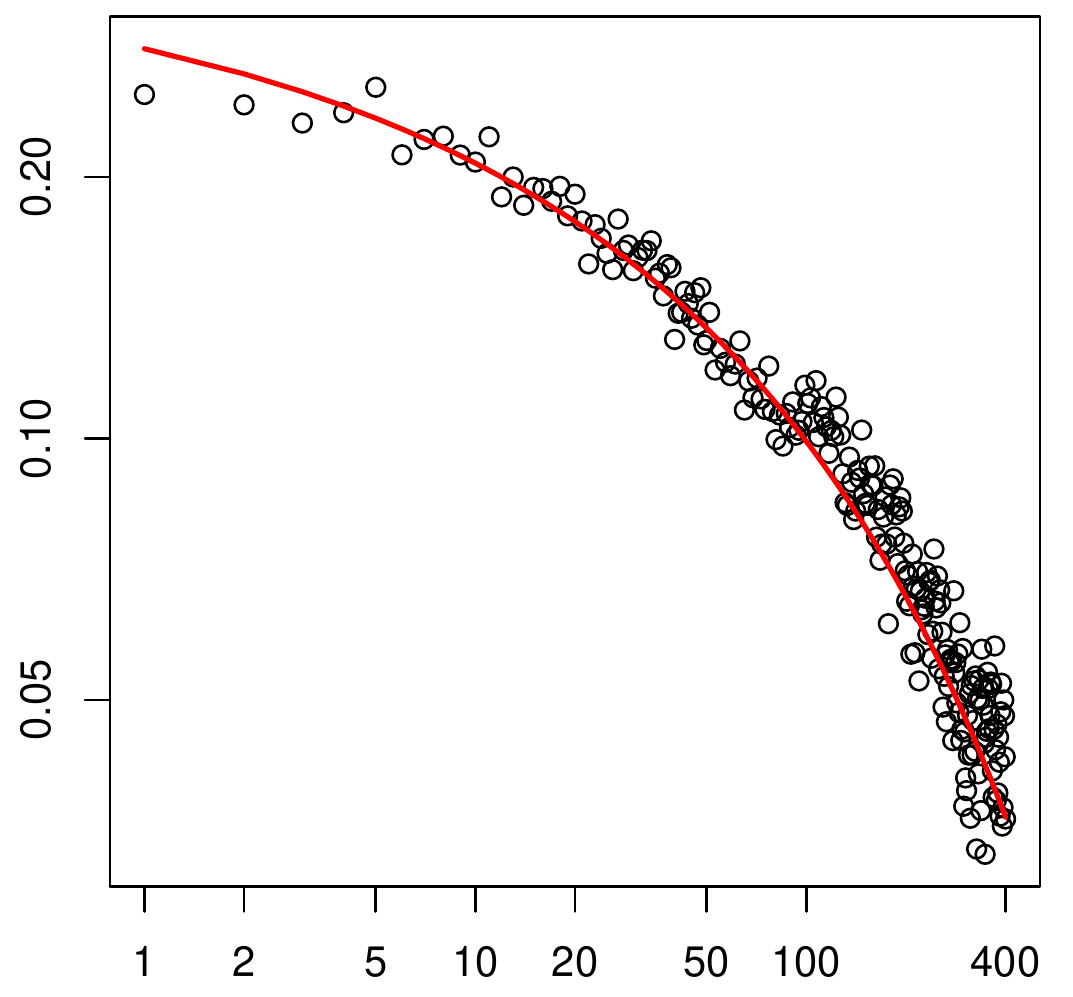}\label{fig:compdji_corr400loglog}}
\caption{\emph{Multiscaling of moments and volatility autocorrelation in the DJIA time series (1935-2009), compared with our model}.\\
{\sc\subref{fig:compmultiscaling2}}
The DJIA empirical scaling exponent $\widehat A(q)$
(circles) and the theoretical scaling exponent $A(q)$ (line)
as a function of $q$. \\
{\sc\subref{fig:compdji_corr400log}}
Log plot for the DJIA empirical 1-day volatility autocorrelation $\widehat \rho_1(t)$
(circles) and the theoretical prediction $\rho(t)$ (line),
as functions of $t$ (days). For clarity, only one data out
of two is plotted.\\
{\sc\subref{fig:compdji_corr400loglog}} Same as {\sc\subref{fig:compdji_corr400log}},
but log-log plot instead of log plot. For clarity, for $t \ge 50$ only one data out
of two is plotted.}
\label{fig:comparisons}
\end{figure}

\begin{figure}
\centering
\subfloat[][\emph{Density of the empirical distribution of the DJIA log-returns
(1935-2009)}.]
{\includegraphics[width=.43\columnwidth]{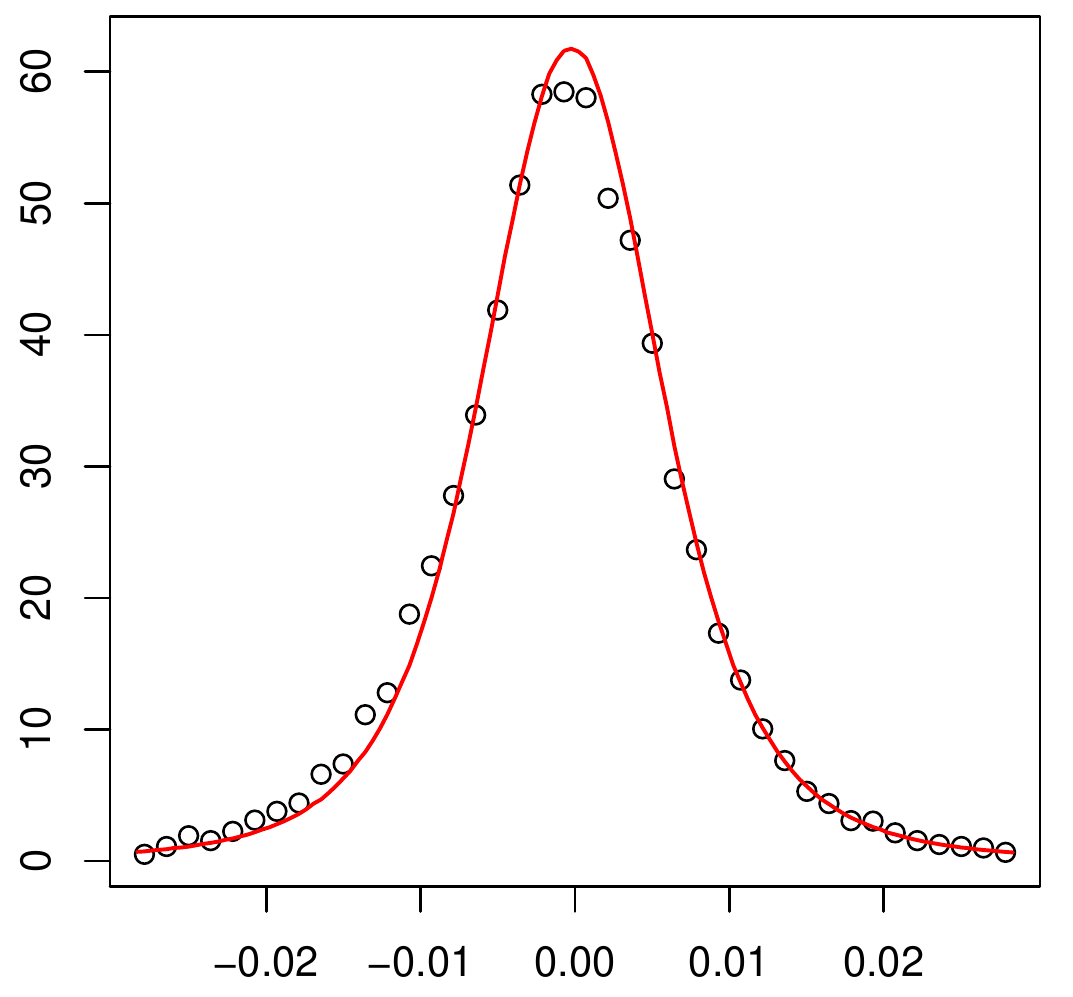}\label{fig:dji_distr}}
\qquad \
\subfloat[][\emph{Tails of the empirical distribution of the DJIA log-returns
(1935-2009)}.]
{\includegraphics[width=.43\columnwidth]{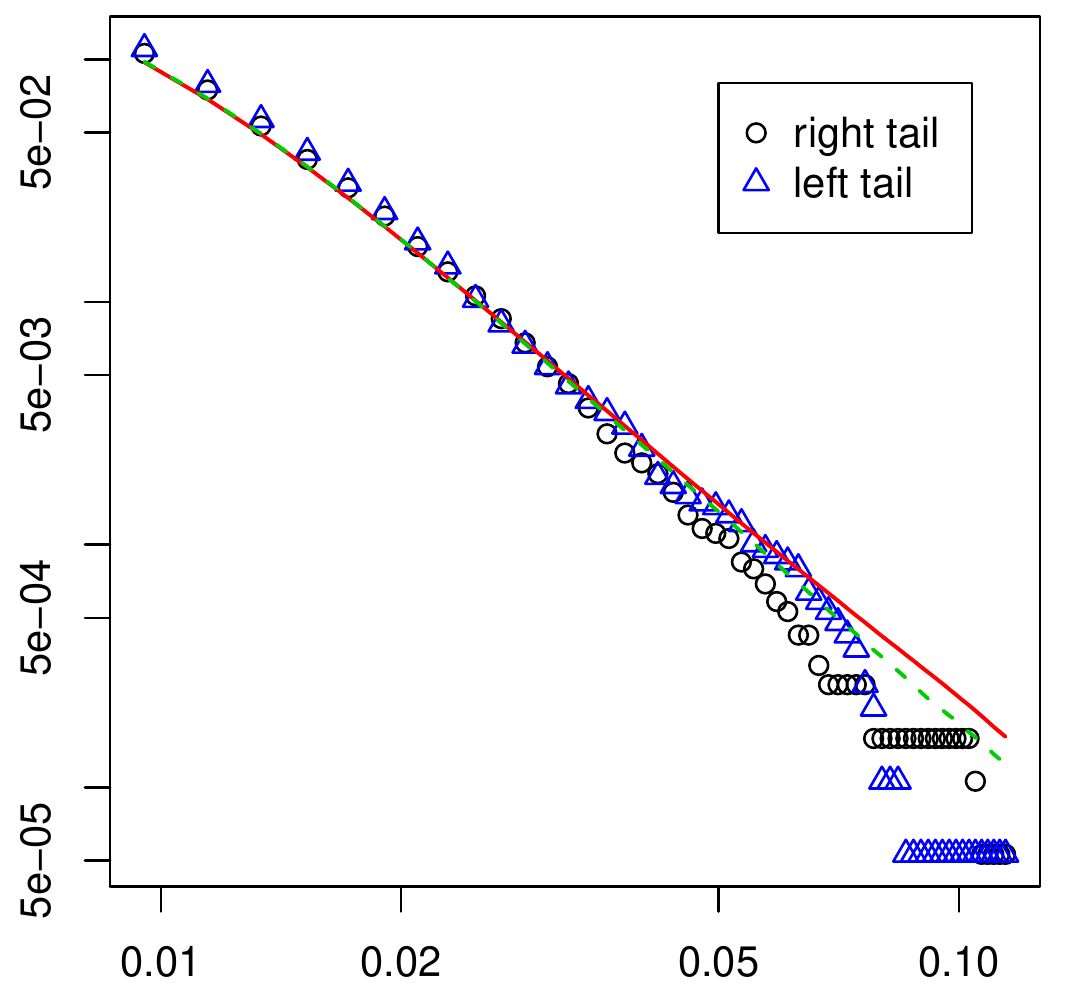}\label{fig:dji_tails}}
\caption{\emph{Distribution of daily log-returns in the DJIA time series (1935-2009),
compared with our model (cf. subsection~\ref{sec:graphcomp} for details)}.\\
{\sc\subref{fig:dji_distr}}
The density of the DJIA log-return empirical distribution $\hat p_1(\cdot)$
(circles) and the theoretical prediction $p_1(\cdot)$ (line). 
The plot ranges from zero to about three standard deviations
($\hat s \simeq 0.0095$) from the mean.\\
{\sc\subref{fig:dji_tails}}
Log-log plot of 
the right and left integrated tails of the DJIA log-return empirical distribution $\hat p_1(\cdot)$
(circles and triangles)
and of the theoretical prediction $p_1(\cdot)$ (solid line). The plot ranges from 
one to about twelve standard deviations from the mean.
Also plotted is the asymptotic density $f(\cdot)$ (dashed line) defined in
equation \eqref{eq:convdist}.}
\label{fig:distribution}
\end{figure}

We now consider some aspects of our model from a numerical viewpoint.
More precisely, we have compared the theoretical predictions
and the simulated data of our model with the time series of some of the
main financial indexes (DJIA, S\&P~500, FTSE~100 and Nikkei~225), finding
a very good agreement. Here we describe in detail the case of the
DJIA time series over a period of 75 years: 
we have considered the DJIA opening prices from 2 Jan 1935 to 31 Dec 2009,
for a total of 18849 daily data.

The four real parameters $D, \lambda, E(\sigma), E(\sigma^2)$ of our model
have been chosen to optimize the fitting of the scaling function $A(q)$ 
of the moments (see Corollary~\ref{A(q)}),
which only depends on $D$, and the curve $\rho(t)$ of the volatility autocorrelation 
(see Corollary~\ref{cor:volauto}), which depends on $D, \lambda, E(\sigma), E(\sigma^2)$
(more details on the parameter estimation are illustrated
in section \ref{sec:estimation}). We have obtained the following numerical estimates:
\begin{equation} \label{eq:parameters}
	\widehat D \simeq 0.16,\quad \widehat \lambda\simeq 0.00097,\quad
	\widehat{E(\sigma)}\simeq 0.108, \quad
	\widehat{E(\sigma^2)} \simeq 0.0117 \simeq \big(\widehat{E(\sigma)} \big)^2 \,.
\end{equation}
Note that the estimated standard deviation of $\s$ is negligible, so
that $\s$ is ``nearly constant''. We point out that the same is true for
the other financial indexes that we have tested. In particular,
in these cases there is no need to specify
other details of the distribution $\nu$ of $\sigma$
and our model is completely determined by the numerical values in \eqref{eq:parameters}.

As we show in Figure~\ref{fig:comparisons}, there is an excellent fitting of the theoretical predictions 
to the observed data. We find remarkable that a rather simple mechanism of 
(relatively rare) volatilty shocks can account for the nontrivial profile 
of both the multiscaling exponent $A(q)$, cf. 
Figure~\ref{fig:comparisons}{\sc\subref{fig:compmultiscaling2}},
and the volatility autocorrelation $\rho(t)$, cf.
Figure~\ref{fig:comparisons}{\sc\subref{fig:compdji_corr400log}}-{\sc\subref{fig:compdji_corr400loglog}}.

\smallskip

Last but not least, we have considered the distribution of daily log-returns:
Figure~\ref{fig:distribution} compares both the density and the integrated tails
of the log-return empirical distribution, cf. \eqref{eq1}, with the theoretical predictions
of our model, i.e., the law of $X_1$. The agreement is remarkable, especially
because the empirical distributions of log-returns \emph{was not used 
for the calibration the model}. This accuracy can
therefore be regarded as structural property of the model. 

In Figure~\ref{fig:distribution}{\sc\subref{fig:dji_tails}} we have plotted
the density of $X_1$, represented by the red solid line, and the asymptotic limiting density $f$
appearing in equation \eqref{eq:convdist} of Theorem~\ref{th:scaling},
represented by the green dashed line.
The two functions are practically indistinguishable up to 
six standard deviations from the mean, and still very close in the whole plotted range.
We stress that $f$ is a rather explicit function, cf. equation \eqref{eq:convdist}.
Also note that the log-log plot in Figure~\ref{fig:distribution}{\sc\subref{fig:dji_tails}}
shows a clear power-law decay, as one would expect from $f$ (the eventually
light tails of $X_1$ are invisible).

\begin{remark}\label{rem:robustness}
We point out that, even if we had found 
$\widehat{Var}(\sigma) := \widehat{E(\sigma^2)} - (\widehat{E(\sigma)})^2 > 0$
(as could happen for different assets),
detailed properties of the distribution of $\sigma$ are not expected
to be detectable from data --- nor are they relevant.
Indeed, the estimated mean distance between the successive epochs
$(\tau_n)_{n\ge 0}$ of the Poisson process $\cT$ is $1/\widehat\lambda \simeq 1031$ days,
cf. \eqref{eq:parameters}.
Therefore, in a time period of the length of the DJIA time series
we are considering, only $18849/1031 \simeq 18$ variables $\sigma_k$ are
expected to be sampled, which is certainly not enough
to allow more than a rough estimation of the distribution of $\sigma$.
This should be viewed more as a robustness than a limitation of our model:
even when $\sigma$ is non-constant, its first two moments
contain the information which is relevant for application to real data.
\end{remark}

\medskip
\section{Discussion and further developments}
\label{sec:discussion}

Now that we have stated the main properties of the model, we can discuss more
in depth its strength as well as its limitations, and consider possible generalizations.

\subsection{On the role of parameters}

A key feature of our model is its rigid structure.
Let us focus for simplicity on the case in which 
\emph{$\sigma$ is a constant} (which, as we have discussed,
is relevant for financial indexes). Not only is
the model characterized by just three real parameters $D, \lambda, \sigma$:
the role of $\lambda$ and $\sigma$ is reduced to \emph{simple scale factors}.
In fact, if we change the value of $\lambda$ and $\sigma$ in our process $(X_t)_{t\ge 0}$,
keeping $D$ fixed, the new process has the same distribution as $(aX_{bt})_{t\ge 0}$ for
suitable $a,b$ (depending of $\lambda,\sigma$), as it is clear from
the definition \eqref{eq10} of $(I_t)_{t\ge 0}$. This means that \emph{$D$ is the only
parameter that truly changes the shape} (beyond simple scale factors)
of the relevant quantities of our model,
as it is also clear from the explicit expressions we have derived for the
small-time and large-time asymptotic distribution (Theorem~\ref{th:scaling}),
mulstiscaling of moments (Theorem~\ref{th:multi}) and volatility
autocorrelation (Theorem~\ref{th:cov}).

More concretely, the structure of our model imposes strict relations
between different quantities: for instance, the moment $q^* = (\frac{1}{2}-D)^{-1}$
beyond which one observes anomalous scaling, cf. \eqref{eq:A(q)},
coincides with the power-law tail exponent of the (approximate) log-return distribution
for small time, cf. \eqref{eq:fqstar}, and is also linked (through $D$) to the slow decay of
the volatility autocorrelation from short to moderate time of the, 
cf. \eqref{eq:phi2} and \eqref{eq:phi3}.
The fact that these quantitative constraints are indeed observed on
the DJIA time series, cf. Figures~\ref{fig:comparisons} and~\ref{fig:distribution},
is not obvious a priori and is therefore particularly noteworthy.

\subsection{On the comparison with multifractal models}

As we observed in the introduction, the multiscaling of moments is a key feature
of multifractal models. These are random time-changes of Brownian motion
$X_t = W_{I_t}$, like our model, with the important difference
that the time-change process $(I_t)_{t\ge 0}$ is rather singular,
having non absolutely continuous paths. Since in our case the time-change process
is quite regular and explicit, our model can be analyzed with more standard
and elementary tools and is very easy to simulate.

A key property of multifractal models,
which is at the origin of their multiscaling features, is that the law of $X_t$ has
power-law tails, for every $t > 0$. On the other hand,
as we already discussed, the law of $X_t$ in our model has
finite moments of all orders --- at least when $E(\sigma^q) < \infty$ for every $q > 0$, which is
the typical case. In a sense, the source of multiscaling in our model
is analogous, because
(approximate) power-law tails appear in the distribution of $X_t$ in the limit $t \downarrow 0$,
but the point is that ``true'' power-law tails in the distribution
of $X_t$ are not necessary to have multiscaling properties.

We remark that the multiscaling exponent $A(q)$
of our model is piecewise linear with two different slopes, 
thus describing a {\em biscaling} phenomenon.
Multifractal models are very flexible in this respect,
allowing for a much wider class of behavior of $A(q)$.
It appears however that a biscaling exponent is compatible 
with the time series of financial indexes (cf.
also Remark~\ref{rem:critics} below).

We conclude with a semi-heuristic argument,
which illustrates how heavy tails and multiscaling arise in our model.
On the event $\{(-\tau_0) \le h, \, \tau_1 > h\}$ we can write, by \eqref{eq10},
$I_h = \sigma_0^2 \{(h- \tau_0)^{2D} - (-\tau_0)^{2D}\} \gtrsim h^{2D}$
and therefore $|X_h| = |W_{I_h}| \sim \sqrt{I_h} \, |W_1| \gtrsim h^D$.
Consequently we get the bound
\begin{equation}\label{eq:lbXh}
	P(|X_h| \gtrsim h^D) \,\ge\, P((-\tau_0) \,\le\, h, \, \tau_1 > h) \gtrsim h \,,
\end{equation}
which allows to draw a couple of interesting consequences.
\begin{itemize}
\item Relation \eqref{eq:lbXh} yields the lower bound
$E(|X_h|^q) \gtrsim h^{Dq} P(|X_h| \gtrsim h^D) \gtrsim h^{Dq+1}$ on the moments of our process.
Since $Dq+1 < q/2$ for $q > q^* = (\frac{1}{2} - D)^{-1}$, this shows
that the usual scaling $E(|X_h|^q) \simeq h^{q/2}$ cannot hold for $q > q^*$.

\item Relation \eqref{eq:lbXh}
can be rewritten as $P(\frac{1}{\sqrt{h}} |X_h| \gtrsim t) \gtrsim t^{-q^*}$,
where $t = h^{-(\frac{1}{2} - D)}$ and $q^* = (\frac{1}{2} - D)^{-1}$. Since
$t \to +\infty$ as $h \downarrow 0$, when $D < \frac{1}{2}$,
this provides a glimpse of the appearance of power law tails 
in the distribution of $X_h$ as $h \downarrow 0$, cf. \eqref{eq:convdist} and \eqref{eq:fqstar},
\emph{with the correct tail exponent $q^*$}.
\end{itemize}

\subsection{On the stochastic volatility model representation}

We recall that our process $(X_t)_{t\ge 0}$ can be written as a stochastic volatility
model $\dd X_t = v_t \, \dd B_t$, cf. \eqref{stochvol}.
It is interesting to note that the squared volatility $(v_t)^2$
is the stationary solution of the following stochastic differential equation:
\be{OU}
\dd (v_t^2) \,=\, - \a_t \left(v_t^2\right)^{\g} \dd t \,+\, \infty \, \dd i(t),
\ee
where we recall that $(i(t))_{t\ge 0}$ is an ordinary Poisson process,
while $\gamma$ is a constant and $\alpha_t$
is a piecewise-constant function, defined by
\[
\g \,:=\, 2+\frac{2D}{1-2D} \,>\,2 \,, \qquad
\a_t \,:=\, \frac{1-2D}{(2D)^{1/(1-2D)}} \frac{1}{\s_{i(t)}^{2/(1-2D)}} \,>\, 0 \,.
\]
We stress that $(v_t)^2$ is a \emph{pathwise}
solution of equation \eqref{OU}, i.e., it solves the equation for any fixed realization
of the stochastic processes $i(t)$ and $\alpha_t$.
The infinite coefficient of the driving Poisson noise is no problem:
in fact, thanks to the {\em superlinear}
drift term $- \a_t \left(v_t^2\right)^{\g}$,
the solution starting from infinity becomes instantaneously 
finite.\footnote{Note that the ordinary differential 
equation $\dd x(t) = - \alpha\, x(t)^\gamma \dd t$
with $x(0) = \infty$ has the explicit solution $x(t) = c \, t^{-1/(\gamma - 1)}$ where
$c = c(\alpha, \gamma) = (\alpha(\gamma - 1))^{-1/(\gamma - 1)}$.}

The representation \eqref{OU} of the volatility is also useful to understand the limitations 
of our model and to design possible generalizations. 
For instance, according to \eqref{OU}, the volatility has the rather unrealistic feature of 
being deterministic between jumps. This limitation could be weakened in various way, e.g. 
by replacing $i(t)$ in \eqref{OU} with a more general Levy subordinator, and/or adding to 
the volatility a continuous random component. Such addition should allow
a more accurate description of the intermittent structure of the volatility profile,
in the spirit of multifractal models.

In a sense, the model we have presented describes only the {\em relatively
rare big jumps} 
of the volatility, ignoring the smaller random fluctuations that are present on smaller time scales.
Besides obvious simplicity considerations, one of our aims is to point out that these big jumps, 
together with a nonlinear drift term as in \eqref{OU},
are sufficient to explain in a rather accurate way the several stylized facts we have discussed.

\subsection{On the skewness and leverage effect}

Our model predicts an even distribution for $X_t$,
but it is known that several financial assets data exhibit a nonzero skewness.
A reasonable way to introduce skewness is through the so-called {\em leverage effect}.
This can be achieved, e.g., by modifying the 
stochastic volatility representation, given in equations \eqref{stochvol} and \eqref{OU}, as follows:
\begin{align*}
\dd X_t & \,=\, v_t\, \dd B_t \,-\, \b\, \dd i(t)\\
\dd v_t^2 & \,=\, - \a_t \left(v_t^2\right)^{\g} \dd t \,+\, \infty\, \dd i(t),
\end{align*}
where $\b>0$. In other words, when the volatility jumps (upward), the price jumps downward by an amount $\g$. The effect of this extension of the model is currently under investigation.

\subsection{On further developments}

A bivariate version $((X_t, Y_t))_{t\ge 0}$ of our model, where
the two components are driven by possibly correlated Poisson point processes $\cT^X, \cT^Y$,
has been investigated by P. Pigato in his Master's Thesis~\cite{cf:Pigato}.
The model has been numerically calibrated on the joint time series
of the DJIA and FTSE~100 indexes, finding in particular a very good agreement  for
the \emph{volatility cross-correlation}
between the two indexes: for this quantity, the model predicts \emph{the same decay profile}
as for the individual volatility autocorrelations, a fact which 
is not obvious a priori and is indeed observed on the real data.

We point out that an important ingredient in the numerical analysis
on the bivariate model is a clever algorithm for finding the location of the
relevant \emph{big jumps} in the volatility
(a concept which is of course not trivially defined). Such an algorithm has
been devised by M. Bonino in his Master's Thesis~\cite{cf:Bonino},
which deals with portfolio optimization problems in the framework of our model.


\medskip
\section{Scaling and multiscaling: proof of Theorems~\ref{th:scaling} and~\ref{th:multi}}
\label{sec:multi}

We observe that for all fixed $t,h > 0$ we have the equality in law
$X_{t+h} - X_t \sim \sqrt{I_h} \, W_1$, as it follows by the definition of our model
$(X_t)_{t\ge 0} = (W_{I_t})_{t\ge 0}$. We also observe
that $i(h) = \#\{\cT \cap (0,h]\} \sim Po(\lambda h)$, as it follows
from \eqref{eq:it} and the properties of the Poisson process.

\subsection{Proof of Theorem~\ref{th:scaling}}


Since $P(i(h) \ge 1) = 1 - e^{-\lambda h} \to 0$
as $h \downarrow 0$, we may focus on the event 
$\{i(h) = 0\} = \{\cT \cap (0,h] = \emptyset\}$, on which we have
$I_h = \sigma_0^2 ((h-\tau_0)^{2D} - (-\tau_0)^{2D})$, with $-\tau_0 \sim Exp(\lambda)$. 
In particular,
\begin{equation*}
	\lim_{h\downarrow 0} \, \frac{I_h}{h} \, = \, 
	I'(0) \,=\, (2D) \, \sigma_0^2 \, (-\tau_0)^{2D-1}
	\qquad \text{a.s.} \,.
\end{equation*}
Since $X_{t+h} - X_t \sim \sqrt{I_h} \, W_1$, the convergence in distribution
\eqref{eq:convdist} follows:
\begin{equation*}
	\frac{X_{t+h} - X_t}{\sqrt{h}} \overset{d}{\longrightarrow} \sqrt{2D} \, \sigma_0 \,
	(-\tau_0)^{D-1/2} \, W_1 \qquad \text{as } h \downarrow 0 \,.
\end{equation*}

Next we focus on the case $h\uparrow \infty$. Under the assumption
$E(\sigma^2) < \infty$, the random variables $\{\sigma_{k-1}^2 (\tau_k - \tau_{k-1})^{2D}\}_{k\ge 1}$
are independent and identically distributed with finite mean, hence by the strong law
of large numbers
\begin{equation*}
	\lim_{n\to\infty} \, \frac{1}{n} \sum_{k=1}^n \sigma_{k-1}^2 (\tau_k - \tau_{k-1})^{2D}
	\,=\, E(\sigma^2) E((\tau_1)^{2D}) \,=\, E(\sigma^2) \, \lambda^{-2D}
	\, \Gamma(2D+1) \qquad \text{a.s.}\,.
\end{equation*}
Plainly, $\lim_{h\to+\infty} i(h)/h = \lambda$ a.s., by the strong law of large
numbers applied to the random variables $\{\tau_k\}_{k\ge 1}$. Recalling
\eqref{eq10}, it follows easily that
\begin{equation*}
	\lim_{h\uparrow \infty} \frac{I(h)}{h} \,=\, E(\sigma^2) \, \lambda^{1-2D}
	\, \Gamma(2D+1) \qquad \text{a.s.}\,.
\end{equation*}
Since $X_{t+h} - X_t \sim \sqrt{I_h} \, W_1$, we obtain the convergence in distribution
\begin{equation*}
	\frac{X_{t+h} - X_t}{\sqrt{h}} \overset{d}{\longrightarrow}
	\sqrt{E(\sigma^2) \, \lambda^{1-2D} \, \Gamma(2D+1)} \, W_1 \qquad
	\text{as } h \uparrow \infty\,,
\end{equation*}
which coincides with \eqref{eq:convdist2}.\qed

\subsection{Proof of Theorem \ref{th:multi}}

Since $X_{t+h} - X_t \sim \sqrt{I_h} \, W_1$, we can write
\begin{equation} \label{eq:proofstart}
	E(|X_{t+h} - X_t|^q) = E(|I_h|^{q/2} |W_1|^q) = 
	E(|W_1|^q) \, E(|I_h|^{q/2}) = c_q \, E(|I_h|^{q/2}) \,,
\end{equation}
where we set $c_q := E(|W_1|^q)$. We therefore focus on $E(|I_h|^{\frac q 2})$,
that we write as the sum of three terms, that will be analyzed separately:
\begin{equation} \label{eq:3terms}
	E(|I_h|^{\frac q 2}) = E(|I_h|^{\frac q 2}
	\, \ind_{\{i(h)=0\}}) + 
	E(|I_h|^{\frac q 2} \, \ind_{\{i(h)=1\}}) 
	+ E(|I_h|^{\frac q 2}\, \ind_{\{i(h)\ge 2\}}) \,.
\end{equation}

For the first term in the right hand side of \eqref{eq:3terms}, 
we note that $P(i(h)=0) = e^{-\lambda h} \to 1$ as
$h \downarrow 0$ and that $I_h =\s_0^2((h-\tau_0)^{2D} - (-\tau_0)^{2D})$
on the event $\{i(h)=0\}$. Setting
$-\tau_0 =: \lambda^{-1} S$ with $S \sim Exp(1)$,
we obtain as $h\downarrow 0$
\begin{equation} \label{eq:term1a}
	E(|I_h|^{\frac q 2}\, \ind_{\{i(h)=0\}}) = E(\s^q)
	\, \lambda^{-Dq} \,
	E\big( ((S+\l h)^{2D} - S^{2D})^{\frac q2} \big) \,
	\big(1+o(1)\big) \,.
\end{equation}
Recalling that $q^* := (\frac 12 - D)^{-1}$, we have
\begin{equation*}
	q \gtreqqless q^* \quad \iff \quad \frac q2 \gtreqqless Dq + 1
	\quad \iff \quad -1 \gtreqqless \left(D - \frac 12 \right) q \,.
\end{equation*}
As $\delta \downarrow 0$ we have $\delta ^{-1}((S+\delta)^{2D} - S^{2D}) \uparrow 
2D\, S^{2D-1}$ and note that $E \big( S^{(D - \frac 12) q} \big)
= \Gamma(1-q/q^*)$ is finite
if and only if $(D - \frac 12)q > -1$, that is $q < q^*$. Therefore
the monotone convergence theorem yields
\begin{equation} \label{eq:term1b}
	\text{for } q < q^*: \quad
	\lim_{h\downarrow 0} 
	\frac{E\big( \big( (S+\l h)^{2D} - S^{2D} \big)^{\frac q2} \big)}
	{\l^{\frac q2} \, h^{\frac q2}}
	\;=\; (2D)^{q/2}\, \Gamma(1-q/q^*) \;\in\; (0,\infty) \,.
\end{equation}
Next observe that, by the change of variables $s = (\l h)x$, we can write
\begin{equation} \label{eq:2l}
\begin{split}
	E\big( ((S+\l h)^{2D} - S^{2D})^{\frac q2} \big) & =
	\int_0^\infty ((s+ \l h)^{2D} - s^{2D})^{\frac q2} \, e^{- s} \, \dd s \\
	& =  (\l h)^{Dq+1} \int_0^\infty ((1+x)^{2D} - x^{2D})^{\frac q2} 
	\, e^{- \l h x} \, \dd x \,.
\end{split}
\end{equation}
Note that $((1+x)^{2D} - x^{2D})^{\frac q2} \sim (2D)^{\frac q2} x^{(D - \frac 12) q}$
as $x \to +\infty$ and that $(D - \frac 12) q < -1$ if and only if $q > q^*$.
Therefore, again by the monotone convergence theorem, we obtain
\begin{equation} \label{eq:term1c}
	\text{for } q > q^*: \quad
	\lim_{h\downarrow 0} \frac{E\big( 
	((S+\l h)^{2D} - S^{2D})^{\frac q2} \big)}
	{\lambda^{Dq+1} \, h^{Dq+1}}
	\;=\; \int_0^\infty ((1+x)^{2D} - x^{2D})^{\frac q2}\, \dd x
	\;\in\; (0,\infty) \,.
\end{equation}
Finally, in the case $q=q^*$ we have
$((1+x)^{2D} - x^{2D})^{q^*/2} \sim (2D)^{q^*/2}\, x^{-1}$
as $x \to +\infty$ and we want to study the integral in the second
line of \eqref{eq:2l}.
Fix an arbitrary (large) $M > 0$ and note that, 
integrating by parts
and performing a change of variables, as $h \downarrow 0$ we have
\begin{align*}
	\int_M^\infty \frac{e^{-\lambda h x}}{x} & \, \dd x \,=\,
	- \log M e^{-\lambda h M} \,+\, \lambda h \int_M^\infty (\log x)\, e^{-\lambda h x} \, \dd x
	\,=\, O(1) + \int_{\lambda h M}^\infty \log \left( \frac{y}{\lambda h} \right)
	e^{-y} \, \dd y \\
	& \,=\, O(1) + \int_{\lambda h M}^\infty \log \left( \frac{y}{\lambda} \right) e^{-y} \, \dd y 
	\,+\, \log \left( \frac{1}{h} \right) \int_{\lambda h M}^\infty	e^{-y} \, \dd y
	\;=\; \log \left( \frac{1}{h} \right) \, \big( 1 + o(1) \big) \,.
\end{align*}
From this it is easy to see that as $h \downarrow 0$
\begin{equation*}
	\int_0^\infty ((1+x)^{2D} - x^{2D})^{\frac{q^*}{2}} 
	\, e^{-\lambda h x} \, \dd x \;\sim\;
	(2D)^{\frac{q^*}{2}} \log \left( \frac{1}{h} \right) \,.
\end{equation*}
Coming back to \eqref{eq:2l}, noting that $Dq+1 = \frac q2$ for $q=q^*$, it follows that
\begin{equation} \label{eq:term1d}
	\lim_{h\downarrow 0} \frac{E\big( ((S+h)^{2D} - S^{2D})^{\frac{q^*}{2}} \big)}
	{\l^{Dq^* + 1} \, h^{\frac{q^*}{2}} \, \log(\frac 1h)}
	= (2D)^{\frac{q^*}{2}} \,.
\end{equation}
Recalling \eqref{eq:proofstart} and \eqref{eq:term1a}, the relations
\eqref{eq:term1b}, \eqref{eq:term1c} and \eqref{eq:term1d} show that the first term
in the right hand side of \eqref{eq:3terms} has the same asymptotic behavior
as in the statement of the theorem, except for the regime $q > q^*$ where
the constant does not match (the missing contribution 
will be obtained in a moment).

We now focus on the second term in the right hand side of \eqref{eq:3terms}.
Note that, conditionally on the event $\{i(h)=1\}
= \{\tau_1 \le h, \tau_2 > h\}$, we have
\begin{equation*}
	I_h = \s_1^2 (h-\tau_1)^{2D} +
	\s_0^2\big( (\tau_1 - \tau_0)^{2D} - (-\tau_0)^{2D} \big)
	\sim \s_1^2 (h-h U)^{2D} + 
	\s_0^2\bigg( \bigg(h U + \frac{S}{\l}\bigg)^{2D} - 
	\bigg(\frac{S}{\l}\bigg)^{2D} \bigg) \,,
\end{equation*}
where $S \sim Exp(1)$ and $U \sim U(0,1)$ (uniformly
distributed on the interval $(0,1)$) are independent of $\s_0$
and $\s_1$.
Since $P(i(h)=1) = \lambda h + o(h)$ as $h \downarrow 0$,
we obtain
\begin{equation} \label{eq:term2a}
	E(|I_h|^{\frac q 2}\, \ind_{\{i(h)=1\}}) = \lambda h^{Dq+1}	\, 
	E \bigg[ \bigg( \s_1^2 (1- U)^{2D} + 
	\s_0^2 \bigg( \bigg( U + \frac{S}{\l h} \bigg)^{2D} - 
	\bigg( \frac{S}{\l h} \bigg)^{2D} \bigg)^{\frac q2} \bigg) \bigg] \,.
\end{equation}
Since $(u + x)^{2D} - x^{2D} \to 0$ as $x \to \infty$,
for every $u \ge 0$, by the dominated convergence theorem we have
(for every $q \in (0,\infty)$)
\begin{equation} \label{eq:term2b}
	\lim_{h\downarrow 0} \frac{E(|I_h|^{\frac q 2}
	\, \ind_{\{i(h)=1\}})}{h^{Dq+1}}
	\;=\; \lambda E(\s_1^q) \, E\big( (1-U)^{Dq} \big) \;=\;
	\lambda E(\s_1^q) \, \frac{1}{Dq+1} \,.
\end{equation}
This shows that the second term in the right hand side of \eqref{eq:3terms}
gives a contribution of the order $h^{Dq+1}$ as $h\downarrow 0$. This is
relevant only for $q>q^*$, because for $q \le q^*$ the first term gives a much bigger
contribution of the order $h^{q/2}$ (see \eqref{eq:term1b} and \eqref{eq:term1d}).
Recalling \eqref{eq:proofstart}, it follows from \eqref{eq:term2b}
and \eqref{eq:term1c} that the contribution of the first and the second term 
in the right hand side of \eqref{eq:3terms} matches the statement
of the theorem (including the constant).

It only remains to show that the third term in the right hand side of \eqref{eq:3terms}
gives a negligible contribution. We begin by deriving a simple upper bound for $I_h$. 
Since $(a+b)^{2D} - b^{2D} \le a^{2D}$ for all $a,b \ge 0$ (we recall that $2D \le 1$),
when $i(h) \ge 1$, i.e. $\tau_1 \le h$, we can write
\begin{equation} \label{ub}
\begin{split}
	I_h & = \s_{i(h)}^2 (h-\tau_{i(h)})^{2D} + \sum_{k=2}^{i(h)} \s_{k-1}^2 
	(\tau_k - \tau_{k-1})^{2D} + \s_0^2 \left[(\tau_1 - \tau_0)^{2D} 
	- (-\tau_0)^{2D} \right] \\
	& \leq \s_{i(h)}^2 (h-\tau_{i(h)})^{2D} + \sum_{k=2}^{i(h)} \s_{k-1}^2
	(\tau_k - \tau_{k-1})^{2D} + \s_0^2 \tau_1^{2D} \,,
\end{split}
\end{equation}
where we agree that the sum over $k$ is zero if $i(h) = 1$.
Since $\tau_k \le h$ for all $k \le i(h)$, by the definition \eqref{eq:it}
of $i(h)$, relation \eqref{ub}
yields the bound $I_h \leq h^{2D}  \sum_{k=0}^{i(h)} \s_{k}^2$,
which holds clearly also when $i(h) = 0$. In conclusion, we have shown that
for all $h, q > 0$
\begin{equation} \label{36}
	|I_h|^{q/2} \le h^{Dq} \Bigg( \sum_{k=0}^{i(h)} \sigma_k^2 
	\Bigg)^{q/2}\,.
\end{equation}
Consider first the case
$q > 2$: by Jensen's inequality we have
\begin{equation} \label{37}
	\Bigg( \sum_{k=0}^{i(h)} \sigma_k^2 
	\Bigg)^{q/2} = (i(h)+1)^{q/2} \,
	\Bigg( \frac{1}{i(h)+1}\, \sum_{k=0}^{i(h)} \sigma_k^2 \Bigg)^{q/2}
	\le (i(h)+1)^{q/2-1} \, \sum_{k=0}^{i(h)} \sigma_k^q \,.
\end{equation}
By \eqref{36} and \eqref{37} we obtain
\begin{equation} \label{eq:Ih1}
	E\big( |I_h|^{q/2} \, \ind_{\{i(h) \ge 2\}} \big) \,\le\, h^{Dq} \, E(\sigma^q) \,
	E\big( (i(h)+1)^{q/2} \, \ind_{\{i(h) \ge 2\}} \big) \,.
\end{equation}
A corresponding inequality for $q \leq 2$ is derived from \eqref{36} and the inequality 
$(\sum_{k=1}^\infty x_k)^{q/2} \le \sum_{k=1}^\infty x_k^{q/2}$, which holds for every non-negative sequence
$(x_n)_{n\in\N}$:
\begin{equation} \label{eq:Ih2}
	E\big( |I_h|^{q/2} \, \ind_{\{i(h) \ge 2\}} \big) \,\le\, h^{Dq}  E \Bigg( \sum_{k=0}^{i(h)} \sigma_k^q  \, \ind_{\{i(h) \ge 2\}}
	\Bigg)      \,\le\,h^{Dq} \, E(\sigma^q) \,
	E\big( (i(h)+1) \, \ind_{\{i(h) \ge 2\}} \big) \,.
\end{equation}
For any fixed $a > 0$, by the H\"older inequality with $p=3$ and
$p'=3/2$ we can write for $h\le 1$
\begin{equation} \label{eq:Ih3}
\begin{split}
	& E\big( (i(h)+1)^a \, \ind_{\{i(h) \ge 2\}} \big) \,\le\,
	E\big( (i(h)+1)^{3a} \big)^{1/3} \, P(i(h) \ge 2)^{2/3} \\
	& \qquad \,\le\, E\big( (i(1)+1)^{3a} \big)^{1/3} \, 
	(1-e^{-\lambda h} - e^{-\lambda h} \lambda h)^{2/3} \,\le\, (const.) \, h^{4/3} \,,
\end{split}
\end{equation}
because $E\big( (i(1)+1)^{3a} \big) < \infty$ (recall that $i(h) \sim Po(\lambda)$)
and $(1-e^{-\lambda h} - e^{-\lambda h} \lambda h) \sim \frac 12 (\lambda h)^2$ as
$h\downarrow 0$. Then it follows from \eqref{eq:Ih1} and \eqref{eq:Ih2}  and \eqref{eq:Ih3} that
\begin{equation*}
	E\big( |I_h|^{q/2} \, \ind_{\{i(h) \ge 2\}} \big) \,\le\, (const.') \, h^{Dq  + 4/3} \,.
\end{equation*}
This shows that the contribution
of the third term in the right hand side of \eqref{eq:3terms} is always negligible
with respect to the contribution of the second term (recall \eqref{eq:term2b}).\qed

\medskip
\section{Decay of correlations: proof of Theorem \ref{th:cov}}
\label{sec:corr}


Given a Borel set $I \subseteq \R$,
we let $\cG_{I}$ denote the $\gs$-algebra generated
by the family of random variables $(\tau_k \ind_{\{\tau_k \in I\}},
\gs_k \ind_{\{\tau_k \in I\}})_{k\ge 0}$.
Informally, $\cG_{I}$ may be viewed as the $\gs$-algebra
generated by the variables $\tau_k, \sigma_k$ for
the values of $k$ such that $\tau_k \in I$.
From the basic property of the Poisson process and from the fact
that the variables $(\sigma_k)_{k\ge 0}$ are independent, it follows that
for disjoint Borel sets $I, I'$ the $\sigma$-algebras
$\cG_I$, $\cG_{I'}$ are independent. We set
for short $\cG := \cG_{\R}$, which is by definition
the $\gs$-algebra generated by all the variables
$(\tau_k)_{k\ge 0}$ and $(\sigma_k)_{k\ge 0}$, which coincides
with the $\gs$-algebra generated by the process $(I_t)_{t\ge 0}$.

We have to prove \eqref{eq:cov}.
Plainly, by translation invariance we can set $s=0$
without loss of generality. We also assume that $h < t$.
We start writing
\begin{equation} \label{eq:cov_pr1}
\begin{split}
	& Cov(|X_{h}|,|X_{t+h}-X_{t}|) \\
	& \;=\;
	Cov\big( E\big(|X_{h}| \big|\cG\big) \,,\, 
	E\big(|X_{t+h}-X_{t}| \big|\cG\big) \big) \;+\;
	E\big( Cov\big(|X_{h}|,|X_{t+h}-X_{t}| \big| \cG\big) \big) \,.
\end{split}
\end{equation}
We recall that $X_t = W_{I_t}$ and the process $(I_t)_{t\ge 0}$
is $\cG$-measurable and independent of the process $(W_t)_{t\ge 0}$. It follows that,
conditionally on $(I_t)_{t\ge 0}$, the process $(X_t)_{t\ge 0}$
has independent increments,
hence the second term
in the right hand side of \eqref{eq:cov_pr1} vanishes, because
$Cov(|X_{h}|,|X_{t+h}-X_{t}| | \cG) = 0$ a.s..
For fixed $h$, from the equality in law $X_h = W_{I_h} \sim \sqrt{I_h} \, W_1$
it follows that $E(|X_{h}| |\cG) = c_1 \sqrt{I_h}$, where
$c_1 = E(|W_1|) = \sqrt{2/\pi}$. Analogously
$E(|X_{t+h}-X_{t}| |\cG) = \sqrt{2/\pi}\sqrt{I_{t+h} - I_t}$ and
\eqref{eq:cov_pr1} reduces to
\begin{equation} \label{eq:cov_pr2}
	Cov(|X_{h}|,|X_{t+h}-X_{t}|) \;=\; \frac{2}{\pi} \,
	Cov\big( \sqrt{I_h} , \sqrt{I_{t+h} - I_t} \big) \,.
\end{equation}

Recall the definitions \eqref{eq:it} and \eqref{eq10} of the variabes $i(t)$
and $I_t$. We now claim that we can replace $\sqrt{I_{t+h} - I_t}$ by 
$\sqrt{I_{t+h} - I_t} \, \ind_{\{\cal T \cap (h,t] = \emptyset\}}$
in \eqref{eq:cov_pr2}. In fact from \eqref{eq10} we can write
\begin{align*}
	I_{t+h} - I_t = \s_{i(t+h)}^2 (t+h-\tau_{i(t+h)})^{2D}
	+ \sum_{k=i(t)+1}^{i(t+h)} \s_{k-1}^2 (\tau_k - \tau_{k-1})^{2D}
	- \s_{i(t)}^2 (t-\tau_{i(t)})^{2D} \,,
\end{align*}
where we agree that the sum in the right hand side is zero
if $i(t+h)=i(t)$. This shows that $(I_{t+h} - I_t)$
is a function of the variables $\tau_k, \sigma_k$ with index $i(t) \le k \le i(t+h)$.
Since $\{\cal T \cap (h,t] \ne \emptyset\} = \{\tau_{i(t)} > h\}$,
this means that $\sqrt{I_{t+h} - I_t} \, \ind_{\{\cal T \cap (h,t] \ne \emptyset\}}$
is $\cG_{(h,t+h]}$-measurable, hence independent of $\sqrt{I_h}$,
which is clearly $\cG_{(-\infty,h]}$-measurable. This shows that
$Cov ( \sqrt{I_h} , \sqrt{I_{t+h} - I_t} \, \ind_{\{\cal T \cap (h,t] \ne \emptyset\}} ) = 0$,
therefore from \eqref{eq:cov_pr2} we can write
\begin{equation} \label{eq:cov_pr3}
	Cov(|X_{h}|,|X_{t+h}-X_{t}|) \;=\; \frac{2}{\pi} \,
	Cov\big( \sqrt{I_h} , \sqrt{I_{t+h} - I_t}\,
	\ind_{\{\cal T \cap (h,t] = \emptyset\}} \big) \,.
\end{equation}
Now we decompose this last covariance as follows:
\begin{equation} \label{eq:cov_pr4}
\begin{split}
	& Cov\big( \sqrt{I_h} , \sqrt{I_{t+h} - I_t}\,
	\ind_{\{\cal T \cap (h,t]  = \emptyset\}}\; = 
	\; E\left[\left( \sqrt{I_h} - E \big(\sqrt{I_h}\big)\right) \sqrt{I_{t+h} - I_t}\,
	\ind_{\{\cal T \cap (h,t]  = \emptyset\}} \right] \\
	& \; = \; E\left[\left( \sqrt{I_h} - E \big(\sqrt{I_h}\big)\right) \sqrt{I_{t+h} - I_t}\,
	\ind_{\{\cal T \cap (0,t+h]  = \emptyset\}} \right]  \\
	& \quad \ \; + \; E\left[\left( \sqrt{I_h} - E \big(\sqrt{I_h}\big)\right) \sqrt{I_{t+h} - I_t}\,
	\ind_{\{\cal T \cap (h,t]  = \emptyset\}} \ind_{\{\cal T \cap ([0,h] \cup (t,t+h])  \neq \emptyset\}}\right] 
\end{split}
\end{equation}
We deal separately with the two terms in the r.h.s. of \eqref{eq:cov_pr4}. The first gives the dominant contribution. To see this, observe that, on $\{\cal T \cap (0,t+h]  = \emptyset\}$
\[
I_h \; = \; \s_0^2 \left[(h-\tau_0)^{2D} - (-\tau_0)^{2D}\right]
\]
and
\[
I_{t+h} - I_t \; = \; \s_0^2  \left[(t+h-\tau_0)^{2D} - (t-\tau_0)^{2D}\right].
\]
Since both $\s_0^2 \left[(h-\tau_0)^{2D} - (-\tau_0)^{2D}\right]$ and $\s_0^2  \left[(t+h-\tau_0)^{2D} - (t-\tau_0)^{2D}\right]$ are independent of $\{\cal T \cap (0,t+h]  = \emptyset\}$, we have
\begin{equation} \label{eq:cov_pr5}
\begin{split}
	& E\left[\left( \sqrt{I_h} - E \big( \sqrt{I_h} \big) \right) \sqrt{I_{t+h} - I_t}\,
	\ind_{\{\cal T \cap (0,t+h]  = \emptyset\}} \right] \\
	& \; = \; E\bigg[\left( \s_0 \sqrt{(h-\tau_0)^{2D} - (-\tau_0)^{2D}} 
	- E \big( \sqrt{I_h} \big) \right) 
	\s_0 \sqrt{(t+h-\tau_0)^{2D} - (t-\tau_0)^{2D}}\,
	\ind_{\{\cal T \cap (0,t+h]  = \emptyset\}} \bigg] \\
	& \; = \; e^{-\l(t+h)} E\left[\left( \s_0 \sqrt{(h-\tau_0)^{2D} - (-\tau_0)^{2D}} 
	- E \big( \sqrt{I_h} \big)
	\right) 
	\s_0 \sqrt{(t+h-\tau_0)^{2D} - (t-\tau_0)^{2D}} \right] \\
	& \; = \; e^{-\l(t+h)} \left\{
	Cov\left( \s_0 \sqrt{(h-\tau_0)^{2D} - (-\tau_0)^{2D}}, 
	\s_0 \sqrt{(t+h-\tau_0)^{2D} - (t-\tau_0)^{2D}}\right) \right. \\
	& \qquad \left. \; + \; \left[ E\left(\s_0 \sqrt{(h-\tau_0)^{2D} - (-\tau_0)^{2D}}\right) 
	- E \big( \sqrt{I_h} \big) \right]
	E\left( \s_0 \sqrt{(t+h-\tau_0)^{2D} - (t-\tau_0)^{2D}}\right)  \right\} .
\end{split}
\end{equation}
Since $\delta^{-1}((\delta+x)^{2D}-x^{2D}) \uparrow 2D x^{2D-1}$ as 
$\delta\downarrow 0$, by monotone convergence we obtain
\begin{equation} \label{eq:cov_pr6}
\begin{split}
& 
\lim_{h \ra 0} \frac{1}{h}  Cov\left( \s_0 \sqrt{(h-\tau_0)^{2D} - (-\tau_0)^{2D}}, \s_0 \sqrt{(t+h-\tau_0)^{2D} - (t-\tau_0)^{2D}}\right) \\
& \; = \; 2D Cov\left( \s_0 (-\tau_0)^{D-1/2} ,  \s_0 (t-\tau_0)^{D-1/2} \right)  \\
& \; = \; 2D \l^{1-2D} Cov\left( \s_0 S^{D-1/2} ,  \s_0 (\l t+S)^{D-1/2} \right) 
 \; = \;  2D \l^{1-2D} \phi(\l t),
\end{split}
\end{equation}
with $S := \l (-\tau_0) \sim Exp(1)$ and $\phi$ is defined in \eqref{eq:phi}. Similarly
\begin{equation} \label{eq:cov_pr7}
\lim_{h \ra 0} \frac{1}{\sqrt{h}} E\left( \s_0 \sqrt{(t+h-\tau_0)^{2D} - (t-\tau_0)^{2D}}\right) \; = \; \sqrt{2D} E\left( \s_0 (t-\tau_0)^{D-1/2} \right) < +\infty.
\end{equation}
Therefore, if we show that
\begin{equation} \label{eq:cov_pr8}
\lim_{h \ra 0}E\left( \sqrt{\frac{I_h}{h}}\right) 
\, = \, \sqrt{2D} E\left( \s_0 (-\tau_0)^{D-1/2} \right)
\end{equation}
using \eqref{eq:cov_pr5}, \eqref{eq:cov_pr6}, \eqref{eq:cov_pr7}, we have
\begin{equation} \label{eq:cov_pr9}
\lim_{h \ra 0} \frac{1}{h} E\left[\left( \sqrt{I_h} - E \sqrt{I_h}\right) \sqrt{I_{t+h} - I_t}\,
	\ind_{\{\cal T \cap (0,t+h]  \; = \; \emptyset\}} \right] = 2D \l^{1-2D} \phi(\l t)
\end{equation}
To complete the proof of \eqref{eq:cov_pr9}, we are left to show \eqref{eq:cov_pr8}. 
But this is a nearly immediate consequence of Theorem \ref{th:multi}:
indeed, using \eqref{eq:Cq} and the fact that $q^* > 1$,
\[
E \sqrt{I_h} \, = \, \frac{1}{E\sqrt{|W_1|}} E(|X_h|) \, = \,  \frac{C_{1}}{E|W_1|} \sqrt{h} + o(\sqrt{h}) \, = \,  \sqrt{2D} E\left( \s_0 (-\tau_0)^{D-1/2} \right)\sqrt{h} + o(\sqrt{h}) \,.
\]

The proof is now completed if we show that the second term in \eqref{eq:cov_pr4} is negligible, i.e. it is $o(h)$. By Cauchy-Schwarz inequality
and the simple fact that $\left( \sqrt{I_h} - E \sqrt{I_h}\right)^2 \leq I_h + E(I_h)$
\begin{equation} \label{eq:cov_pr10}
\begin{split}
& E\left[\left( \sqrt{I_h} - E \sqrt{I_h}\right) \sqrt{I_{t+h} - I_t}\,
	\ind_{\{\cal T \cap (h,t]  = \emptyset\}} \ind_{\{\cal T \cap ((0,h] \cup (t,t+h])  \neq \emptyset\}}\right]  \\
	& \; \leq \; \left(E\left[\left( \sqrt{I_h} - E \sqrt{I_h}\right)^2 \left(I_{t+h} - I_t\right) \right] P(\cal T \cap ((0,h] \cup (t,t+h])  \neq \emptyset) \right)^{1/2} \\
	& \; \leq \; \left(\, E\left[(I_h + E(I_h))
	\left(I_{t+h} - I_t\right) \right] P(\cal T \cap ((0,h] \cup (t,t+h])  \neq \emptyset) \right)^{1/2} \\
	&  \; \leq \; \left(2 \, E\left[I_h^2 \right] 
	P(\cal T \cap ((0,h] \cup (t,t+h])  \neq \emptyset) \right)^{1/2}
	\; = \; \left(2 \, E\left[I_h^2 \right]\right)^{1/2} \sqrt{2\l h} \,.
\end{split}
\end{equation}
By Theorem \ref{th:multi}, 
$E\left[I_h^2 \right] $ is of order $h^2$ if $4 < q^*$, and of order $h^{4D+1}$ if $4>q^*$, with a logarithmic correction for $q^* = 4$. In both cases $\left(E\left[I_h^2 \right]\right)^{1/2} \sqrt{2\l h} = o(h)$, and the proof is completed.
\qed 


\medskip
\section{Basic properties of the model}
\label{sec:basic}

In this Section we start by proving the properties 
\eqref{prop:stationarity}-\eqref{prop:momentsXsigma}-\eqref{prop:randomvol}-\eqref{prop:martingale}
stated in section~\ref{subsec:basic}. Then we provide some connections between the tails of $\s$ and those of $X_t$, also beyond the equivalence stated in \eqref{eq:qXsigma}. Finally, we establish a {\em mixing} property that yields relation \eqref{eq:erg}. 
One of the proofs is postponed to the Appendix.

We denote by $\mathcal{G}$ the $\s$-field generated 
by the whole process $(I_t)_{t\ge 0}$,
which coincides with the
$\sigma$-field generated by the sequences $\cT = \{\tau_k\}_{k\ge 0}$
and $\Sigma =\{\sigma_k\}_{k\ge 0}$.


\medskip
\noindent
{\it Proof of property~\eqref{prop:stationarity}.\ }
We first focus on the process $(I_t)_{t\ge 0}$, defined in \eqref{eq10}.
For $h>0$ let ${\cal{T}}^h := {\cal{T}} - h$ and denote the points in $\cT^h$ by 
$\tau_k^h = \tau_k - h$. As before, let 
$\tau^h_{i^h(t)}$ be the largest point of  ${\cal{T}}^h$ smaller that $t$,
i.e., $i^h(t) = i(t+h)$. Recalling the definition \eqref{eq10}, we can write
\[
I_{t+h} - I_h = \s_{i^h(t)}^2 \left(t - \tau^h_{i^h(t)} \right)^{2D} +
\sum_{k= i^h(0)+1}^{i^h(t)} \s_{k-1}^2 \left(\tau^h_{k} - \tau^h_{k-1} \right)^{2D} - 
\s^2_{i^h(0)} \left(-\tau^h_{i^h(0)} \right)^{2D},
\]
where we agree that the sum in the right hand side is zero if $i^h(t)=i^h(0)$.
This relation shows that $(I_{t+h} - I_h)_{t\ge 0}$ and $(I_t)_{t\ge 0}$
are \emph{the same function} of the two random sets ${\cal{T}}^h$ and ${\cal{T}}$.
Since ${\cal{T}}^h$ and ${\cal{T}}$ have the same distribution
(both are Poisson point processes on $\R$ with intensity $\lambda$),
the processes $(I_{t+h} - I_h)_{t \geq 0}$ and $(I_t)_{t \geq 0}$
have the same distribution too. 

We recall that $\mathcal{G}$ is the $\s$-field generated 
by the whole process $(I_t)_{t\ge 0}$.
From the definition $X_t = W_{I_t}$ and
from the fact that Brownian motion has independent, stationary increments,
it follows that for every Borel subset $A \subseteq \R^{[0,+\infty)}$
\begin{equation*}
P\left( X_{h+\cdot} - X_h \in A \right) = 
E\left[ P\left( W_{I_{h+\cdot}} - W_{I_h} \in A \,|\, \mathcal{G} \right) \right] 
= P\left( W_{I_{\cdot}}\in A \right)
= P\left( X_{\cdot} \in A \right),
\end{equation*}
where we have used the stationarity property of the process $I$.
Thus the processes $(X_t)_{t\ge 0}$ and $(X_{h+t} - X_h)_{t\ge 0}$ have the 
same distribution, which implies stationarity of the increments.\qed

\medskip
\noindent
{\em Proof of property \eqref{prop:momentsXsigma}.}
Note that
$E(|X_t|^q) = E(|W_{I_t}|^q) = E(|I_t|^{q/2}) \, E(|W_1|^q)$, 
by the independence of $W$ and $I$ and the scaling properties of Brownian motion.
We are therefore left with showing that
\begin{equation} \label{eq:qXsigma2}
	E(|I_t|^{q/2}) < \infty 
	\quad\Longleftrightarrow\quad E(\sigma^q) < \infty \,.
\end{equation}
The implication ``$\Rightarrow$''
is easy: by the definition \eqref{eq10} of the process $I$ we can write
\begin{equation*}
	E(|I_t|^{q/2}) \ge E(|I_t|^{q/2} \, \ind_{\{i(t) = 0\}})
	= E(\sigma_0^q) \, E(|(t-\tau_0)^{2D} -(-\tau_0)^{2D}|^{q/2}) P(i(t) = 0) \,,
\end{equation*}
therefore if $E(\sigma^q) = \infty$ then also $E(|I_t|^{q/2}) = \infty$.

The implication ``$\Leftarrow$'' follows immediately from the bounds \eqref{eq:Ih1} and \eqref{eq:Ih2}, which hold also without the 
indicator $\ind_{\{i(h) \ge 2\}}$.
\qed

\medskip
\noindent
{\it Proof of property~\eqref{prop:randomvol}.\ }
Observe first that
$I'_s := \frac{\dd}{\dd s} I_s > 0$ a.s. and for Lebesgue--a.e.~$s \ge 0$.
By a change of variable, we can rewrite the process $(B_t)_{t\ge 0}$ defined
in \eqref{volatility} as
\[
	B_t = \int_0^{I_t} \frac{1}{\sqrt{I'(I^{-1}(u))}} \, \dd W_u
	= \int_0^t \frac{1}{\sqrt{I'_s}} \,\dd W_{I_s}
	= \int_0^t \frac{1}{\sqrt{I'_s}} \,\dd X_s \,,
\]
which shows that relation \eqref{stochvol} holds true.
It remains to show that $(B_t)_{t\ge 0}$
is indeed a standard Brownian motion. Note that
\begin{equation*}
	B_t 	= \int_0^{I_t} \sqrt{(I^{-1})'(u)} \, \dd W_u \,.
\end{equation*}
Therefore, conditionally on $\cal G$
(the $\sigma$-field generated by $(I_t)_{t\ge 0}$), $(B_t)_{t\ge 0}$ is a centered Gaussian
process --- it is a Wiener integral --- with conditional covariance given by
\begin{equation*}
	Cov(B_s, B_t \,|\, \cal G) = \int_0^{\min\{I_{s},I_{t}\}} 
	(I^{-1})'(u) \, \dd u = \min\{s,t\} \,.
\end{equation*}
This shows that, conditionally on $\cal G$, $(B_t)_{t\ge 0}$ is a Brownian motion.
Therefore, it is \emph{a fortiori} a Brownian motion without conditioning.\qed

\medskip
\noindent
{\it Proof of property~\eqref{prop:martingale}.\ }
The assumption $E(\gs^2) < \infty$ ensures that $E(|X_t|^2) < \infty$ for all $t \ge 0$,
as we have already shown.
Let us now denote by $\cF^X_t = \gs(X_s, \, s \le t)$ the
natural filtration of the process $X$. We recall that 
$\cG$ denotes the $\gs$-field generated by the whole
process $(I_t)_{t\ge 0}$ and we denote by $\mathcal{F}_t^X \vee \mathcal{G}$
the smallest $\gs$-field containing $\cF^X_t$ and $\mathcal{G}$.
Since $E(W_{I_{t+h}} - W_{I_{t}} | \mathcal{F}_t^X \vee \mathcal{G}) = 0$
for all $h \geq 0$, by the basic properties of Brownian motion,
recalling that $X_t = W_{I_t}$ we obtain
\[
E(X_{t+h} | \mathcal{F}_t^X \vee \mathcal{G}) = 
X_t + E(W_{I_{t+h}} - W_{I_{t}} | \mathcal{F}_t^X \vee \mathcal{G})
= X_t.
\]
Taking the conditional expectation with respect to $\mathcal{F}_t^X$
on both sides, we obtain the martingale property for $(X_t)_{t\ge 0}$.\qed

%

\medskip

Let us state a proposition, proved in Appendix~\ref{sec:appetails},
that relates the exponential moments of $\s$ to those of $X_t$.
We recall that, when our model is calibrated to real time series,
like the DJIA, the ``observable tails'' of $X_t$ are quite insensitive
to the details of the distribution of $\sigma$, cf. Remarks~\ref{rem:realtails}
and~\ref{rem:tails}.

\bp{prop:etails}
Regardless of the distribution of $\gs$, for every $q > (1-D)^{-1}$ we have
\begin{equation} \label{eq:condexpmomA}
	E \left[ \exp\left( \gamma |X_t|^q \right) \right] = \infty \,, \qquad
	\forall t > 0\, , \ \forall \gamma > 0 \,.
\end{equation}
On the other hand, for all $q < (1-D)^{-1}$ and $t > 0$ we have
\begin{equation} \label{eq:condexpmomB}
	E \left[ \exp\left(\gamma |X_t|^q \right) \right] < \infty \quad
	\forall \gamma > 0 \ \quad \ \Longleftrightarrow
	\quad \ E \left[ \exp\left( \alpha \sigma^{\frac{2q}{2-q}} \right) \right] < \infty
	\quad \forall \alpha > 0 \,,
\end{equation}
and the same relation holds for $q=(1-D)^{-1}$ provided $D < \frac 12$.
\ep

\noindent
Note that $(1-D)^{-1} \in (1,2]$, because $D \in (0,\frac 12]$,
so that for $D < \frac 12$ the distribution of $X_t$ has always
tails heavier than Gaussian.

\smallskip

We finally show a mixing property for the increments of our
process. In what follows, for an interval $I \subseteq [0,+\infty)$, we let 
\[
\FF^{\D}_I := \s\left( X_t - X_s : s,t \in I\right)
\]
to denote the $\s$-field generated by the increments in $I$ of the process $X$.
\bp{prop:ergodicity}
Let $I = [a,b)$, $J = [c,d)$, with $0 \leq a < b \leq c < d$. Then, for every $A \in \FF^{\D}_I$ and $B \in \FF^{\D}_J$
\be{mixing}
|P(A \cap B) - P(A)P(B)| \leq e^{- \l(c-b)} \,.
\ee
 As a consequence,
equation \eqref{eq:erg} holds true almost surely and in $L^1$,
for every measurable function $F: \R^k \to \R$ such that 
$E[|F(X_{b_1} - X_{a_1}, \ldots, X_{b_k} - X_{a_k})|] < +\infty$.
\ep

\bpr
We recall that ${\cal{T}}$ denotes the set $\{\tau_k: \ k\in \Z\}$ and, for $I \subseteq \R$, $\cG_{I}$ denotes the $\gs$-algebra generated
by the family of random variables $(\tau_k \ind_{\{\tau_k \in I\}},
\gs_k \ind_{\{\tau_k \in I\}})_{k\ge 0}$,
where $(\sigma_k)_{k\ge 0}$
is the sequence of volatilities. 
We introduce the $\cG_{[b,c)}$-measurable event
\[
\G := \{ {\cal{T}} \cap [b,c) \neq \emptyset \}.
\]
(We recall that the $\sigma$-field $\cG_I$ was defined at the beginning
of section~\ref{sec:corr}.)
We claim that, for $A \in \FF^{\D}_I$, $B \in \FF^{\D}_J$, we have
\be{mixing1}
P(A \cap B\cap \G) = P(A)P(B \cap \G ).
\ee
To see this, the key is in the following two remarks.
\bi
\item
$\FF^{\D}_I$ and $\FF^{\D}_J$ are independent {\em conditionally on $\cG = \cG_{\R}$}. This follows immediately from the independence of $W$ and $(I_t)$. As a consequence,
$P(A \cap B | \cG) = P(A|\cG) P(B|\cG)$ a.s..

\item
Conditionally to $\cG$, the family of random variables $(X_t - X_s)_{s,t \in [c,d)}$ is
a Gaussian process whose covariances are measurable with respect to the $\s$-field 
generated by the random variables $\{I_t - I_c: \, t \in (c,d)\}$.
In particular, $P(B|\cG)$ is measurable with respect to this $\s$-field.
Similarly for $[a,b)$ in place of $[c,d)$.
Note also that the increment $I_t - I_c$ is a measurable function of the random variables 
\[
\{(\tau_k \ind_{\{\tau_k \in I\}},
\gs_k \ind_{\{\tau_k \in I\}}): \, k\ge 0 \} \cup \{(\s_{i(c)}, \tau_{i(c)})\}.
\]
It follows that the random variable
$(P(B|\cG) \ind_{\G})$ is $\cG_{(b,d)}$ measurable, and it is therefore independent of $P(A|\cG)$, which is $\cG_{(-\infty,b]}$ measurable. 
\ei
Thus we have
\[
P(A \cap B \cap \G) = E(P(A \cap B|\cG)\ind_{\G}) = E(P(A|\cG) P(B | \cG)\ind_{\G}) = P(A)P(B \cap \G)
\]
where the two remarks above have been used. Thus \eqref{mixing1} is established. Finally 
\[
\begin{split}
& |P(A \cap B) - P(A) P(B)| \\ & = |P(A \cap B \cap \G) + P(A \cap B \cap \G^c) - P(A)P(B \cap \G) - P(A)P(B \cap \G^c)| \\ & = |P(A \cap B \cap \G^c)- P(A)P(B \cap \G^c)|
= |P(A\cap B|\Gamma^c)-P(A|\Gamma^c)P(B| \Gamma^c)|P(\Gamma^c) \\
& \le P(\Gamma^c) = e^{-\l (c-b)} \,.
\end{split}
\end{equation*}

We finally show that equation \eqref{eq:erg} holds true almost surely and in $L^1$,
for every measurable function $F: \R^k \to \R$ such that 
$E[|F(X_{b_1} - X_{a_1}, \ldots, X_{b_k} - X_{a_k})|] < +\infty$.
Consider the $\R^k$-valued stochastic process
$\xi = (\xi_n)_{n\in\N}$ defined by
\[
\xi_n := (X_{n\d + b_1} - X_{n\d + a_1}, \ldots, X_{n\d + b_k} - X_{n\d + a_k}) \,,
\]
for fixed $\d>0$, $k \in \N$ and
$(a_1, b_1)$, \ldots, $(a_k, b_k) \subseteq (0,\infty)$.
The process $\xi$ is stationary, because we have proven
in section~\ref{sec:basic} that $X$ has stationary increments. Moreover, inequality \eqref{mixing} implies that $\xi$ is {\em mixing}, and therefore ergodic (see e.g. \cite{Sh}, Ch. 5, \S 2, Definition 4 and Theorem 2). 
 The existence of the limit
in \eqref{eq:erg}, both a.s. and in $L^1$, is then a consequence
of the classical Ergodic Theorem, (see e.g. \cite{Sh}, Ch. 5, \S 3, Theorems 1 and 2). 
\epr

\medskip
\section{Estimation and data analysis} \label{sec:estimation}

\label{sec:numerics}

In this Section we present the main steps that led to the calibration of the model to the DJIA over a period of 75 years; the essential results have been sketched in Section \ref{subsec:fitting}.
We point out that the agreement with the S\&P~500, FTSE~100 and Nikkei~225 indexes
is very good as well. A systematic treatment of other time series, beyond
financial indexes, still has to be done, but some preliminary analysis of 
single stocks shows that our model fits well some but not all of them. 
It would be interesting to understand which of the properties we have 
mentioned are linked to \emph{aggregation} of several stock prices, as in the DJIA.

The data analysis, the simulations and the plots have been 
obtained with the software R \cite{R}. 
The code we have used is publicly available on the web
page \url{http://www.matapp.unimib.it/~fcaraven/c.html}.


\subsection{Overview}

For the numerical comparison
of our process $(X_t)_{t\ge 0}$ with the DJIA
time series, we have decided to focus on the following quantities:
\begin{aenumerate}
\item The \emph{multiscaling of moments}, cf. Corollary~\ref{A(q)}.
\item The \emph{volatility autocorrelation decay}, cf. Corollary~\ref{cor:volauto}.
\end{aenumerate}
Roughly speaking, the idea is to compute \emph{empirically} these quantities
on the DJIA time series
and then to compare the results with the \emph{theoretical}
predictions of our model. This is justified
by the ergodic properties of the
increments of our process $(X_t)_{t\ge 0}$,
cf. equation~\eqref{eq:erg}.

\smallskip

The first problem that one faces is the
\emph{estimation of the parameters} of our model:
the two scalars $\l \in (0,\infty)$, $D \in (0,\frac 12]$
and the \emph{distribution $\nu$ of $\s$}. This in principle
belongs to an infinite dimensional space, but in a first time we focus on
the moments $E(\sigma)$ and $E(\sigma^2)$.
In order to estimate $(D,\lambda,E(\sigma),E(\sigma^2))$,
we take into account four significant quantities that depend only on these parameters: 
\begin{itemize}
\item the multiscaling coefficients $C_1$ and $C_2$ (see \eqref{eq:Cq});
\item the multiscaling exponent $A(q)$ (see \eqref{eq:A(q)});
\item the volatility autocorrelation function $\rho(t)$ (see \eqref{eq:volauto}).
\end{itemize}
We consider a natural loss functional $\cL = \cL (D,\lambda,E(\sigma),E(\sigma^2))$
which measures the distance between these theoretical quantities and
the corresponding empirical ones, evaluated on the DJIA time series, see \eqref{eq:L} below.
We then define the estimator for $(D,\lambda,E(\sigma),E(\sigma^2))$
as the point at which $\cL$ attains its 
overall minimum, subject to the constraint
$E(\sigma^2) \geq (E(\sigma))^2$.

It turns out that the estimated values
are such that $E(\sigma^2) \simeq (E(\sigma))^2$,
that is $\sigma$ \emph{is nearly constant} and the estimated parameters completely
specify the model.
(The constraint $E(\sigma^2) \geq (E(\sigma))^2$ is not playing a relevant role: the unconstrained minimum nearly coincides with the constrained one.) Thus, the problem
of determining the distribution of $\sigma$ beyond its moments $E(\sigma)$ and $E(\sigma^2)$
does not appear in the case of the DJIA. More generally, even
we had found $\widehat{Var}(\sigma) := \widehat{E(\sigma^2)} - (\widehat{E(\sigma)})^2 > 0$
and hence $\sigma$ is not constant,
fine details of its distribution $\nu$ beyond the first two moments
give a negligible contribution to the properties that are relevant
for application to real data series, as we observed in Remark~\ref{rem:robustness}.



\subsection{Estimation of the parameters $D,\lambda,E(\sigma),E(\sigma^2)$}

\label{sec:calD}

Let us fix some notation: the DJIA time series will be denoted
by $(s_i)_{0 \le i \le N}$ (where $N = 18848$) and the corresponding
detrended log-DJIA time series will be denoted by $(x_i)_{0 \le i \le N}$:
\begin{equation*}
	x_i := \log(s_i) - \overline {d}(i) \,,
\end{equation*}
where $\overline{d}(i) :=\frac{1}{250}\sum_{k=i-250}^{i-1}{\log(s_i)}$ is
the mean log-DJIA price on the previous 250 days.
(Other reasonable choices for $\overline{d}(i)$ affect
the analysis only in a minor way.)

The theoretical scaling exponent $A(q)$ is defined in \eqref{eq:A(q)}
while the multiscaling constants $C_1$ and $C_2$ are given
by \eqref{eq:Cq} for $q=1$ and $q=2$. Since
$q^* = (\frac{1}{2} - D)^{-1} > 2$ (we recall that $0 \leq D \le \frac{1}{2}$),
we can write more explicitly
\begin{equation} \label{eq:sigma12}
	C_1  = \frac{2 \,
	\sqrt{D} \, \Gamma(\frac 12 + D) \, E(\sigma) \,\lambda^{1/2-D}}{\sqrt{\pi} } \,, \qquad
	C_2  =  2D \,
	\Gamma(2D) \, E(\sigma^2) \, \lambda^{1-2D} \,.
\end{equation}
Defining the corresponding empirical quantities
requires some care, because the DJIA data are in discrete-time
and therefore no $h \downarrow 0$ limit is possible.
We first evaluate the empirical $q$-moment $\widehat m_q(h)$ of the
DJIA log-returns over $h$ days, namely
\begin{equation*}
	\widehat m_q(h) := \frac{1}{N+1-h} \sum_{i=0}^{N-h} | x_{i+h} - x_i |^q \,.
\end{equation*}
By Theorem~\ref{th:multi}, the relation $\log \widehat m_q(h) \sim A(q) (\log h) + \log(C_q)$
should hold for $h$ small. By plotting $(\log \widehat m_q(h))$ versus $(\log h)$ one
finds indeed an approximate linear behavior, for moderate values
of $h$ and when $q$ is not too large ($q \lesssim 5$).
By a standard linear regression of $(\log \widehat m_q(h))$ versus
$(\log h)$ for $h=1,2,3,4,5$ days we therefore determine the
empirical values of $A(q)$ and $C_q$ on the DJIA time series,
that we call $\widehat A(q)$ and $\widehat C_q$.

%

For what concerns the
theoretical volatility autocorrelation, Corollary~\ref{cor:volauto}
and the stationarity of the increments of our process $(X_t)_{t\ge 0}$ yield
\begin{equation} \label{eq:rho}
	\rho(t) \,:=\, \lim_{h\downarrow 0} \rho(|X_{h}|,|X_{t+h}-X_{t}|) \,=\,
	\frac{2}{\pi \, Var(\sigma\, |W_1|\, S^{D-1/2})} \, e^{-\lambda t}
	\, \phi( \lambda t) \,,
\end{equation}
where $S \sim Exp(1)$ is independent of $\sigma$ and $W_1$ and where the function
$\phi(\cdot)$ is given by
\begin{equation*}
	\phi(x) \;=\; Var(\s) \, E(S^{D-1/2}\, (S+x)^{D-1/2}) \,+\,
	E(\s)^2 \, Cov(S^{D-1/2}, (S+x)^{D-1/2}) \,,
\end{equation*}
cf. \eqref{eq:phi2}.
Note that, although $\phi(\cdot)$ does not admit an explicit expression, 
it can be easily evaluated numerically.
For the analogous empirical quantity, we define the
empirical DJIA volatility autocorrelation $\widehat \rho_h(t)$ over $h$-days  
as the \emph{sample correlation coefficient} of the two sequences 
$(|x_{i+h}-x_i|)_{0 \le i \le N-h-t}$ and $(|x_{i+h+t}-x_{i+t}|)_{0 \le i \le N-h-t}$.
Since no $h \downarrow 0$ limit can be taken on discrete data, we are going
to compare $\rho(t)$ with $\widehat \rho_h(t)$ for $h=1$ day.

We can then define a \emph{loss functional $\cL$} as follows:
\begin{equation}\label{eq:L}
\begin{split}
	\cL(D,\lambda,E(\sigma),E(\sigma^2)) \;=\; 
	& \frac 12 \, \bigg\{ \bigg(\frac{\widehat C_1}{C_1}-1\bigg)^2
	\,+\, \bigg(\frac{\widehat C_2}{C_2}-1\bigg)^2 \bigg\}
	\, + \, \frac{1}{20} \sum_{k=1}^{20}\bigg(\frac{\widehat A(k/4)}{A(k/4)}-1 \bigg)^2 \\
	& + \, 	\sum_{n=1}^{400} \frac{e^{-n/T}}{\big(\sum_{m=1}^{400} e^{-m/T} \big)}
	\bigg(\frac{\widehat \rho_1(n)}{\rho(n)}-1 \bigg)^2,
\end{split}
\end{equation}
where the constant $T$ controls a discount factor in long-range correlations.
Of course, different weights for the four terms appearing
in the functional could be assigned.
We fix $T=40$ (days) and we define the estimator
$(\widehat D,\widehat \lambda, \widehat{E(\sigma)}, \widehat{E(\sigma^2)})$ 
of the parameters of our model
as the point where the functional $\cL$ attains its overall minimum, that is
\begin{equation*}
	(\widehat D,\widehat \lambda, \widehat{E(\sigma)}, \widehat{E(\sigma^2)})
	\,:=\, \argmin_{\substack{D \in (0,\frac 12], \
	\lambda , \, E(\sigma), \, E(\sigma^2) \in (0,\infty) \\
	\text{such that} \ E(\sigma^2) \ge (E(\sigma))^2}} 
	\big\{ \cL(D,\lambda,E(\sigma),E(\sigma^2)) \big\} \,,
\end{equation*}
where the constraint $E(\sigma^2) \ge (E(\sigma))^2$ is due to
$Var(\sigma) = E(\sigma^2) - (E(\sigma))^2 \ge 0$. We expect
that such an estimator has good properties, such as asymptotic consistency and
normality (we omit a proof of these properties,
as it goes beyond the spirit of this paper).

We have then proceeded to the numerical study of the functional $\cL$,
which appears to be quite regular. With the help of
the software Mathematica
\cite{cf:Mathematica}, we have obtained the estimates for the parameters, given already in \eqref{eq:parameters}:
\begin{equation}\label{eq:estpar}
	\widehat D \simeq 0.16,\quad \widehat \lambda\simeq 0.00097,\quad
	\widehat{E(\sigma)}\simeq 0.108, \quad
	\widehat{E(\sigma^2)} \simeq 0.0117 \simeq \big(\widehat{E(\sigma)} \big)^2
\end{equation}

\subsection{Graphical comparison}
\label{sec:graphcomp}

Having found that $\widehat{E(\sigma^2)} \simeq
(\widehat{E(\sigma)})^2$, 
the estimated variance of $\sigma$ is equal to zero, that is $\sigma$ is
a constant. In particular, the model is completely specified and we
can compare some quantities, as predicted by our model,
with the corresponding numerical ones evaluated on the DJIA time series.
The graphical results have been already described
in section~\ref{subsec:fitting} and show a very good agreement, 
cf. Figure~\ref{fig:comparisons}
for the multiscaling of moments and the volatility autocorrelation
and Figure~\ref{fig:distribution} for the log-return distribution.

\smallskip

Let us give some details about Figure~\ref{fig:distribution}.
The theoretical distribution 
$p_t(\cdot) := P(X_t \in \cdot) = P(X_t-X_0 \in \cdot)$ of our model,
for which we do not have an analytic expression,
can be easily evaluated numerically via Monte Carlo simulations.
The analogous quantity evaluated for the
DJIA time series is the empirical distribution $\widehat p_t(\cdot)$ of the sequence
$(x_{i+t}-x_i)_{0 \le i \le N-t}$:
\begin{equation}
	\widehat p_t(\cdot) := \frac{1}{N+1-t} \sum_{i=0}^{N-t} \d_{x_{i+t} - x_i}(\cdot) \,.
\end{equation}
In Figure~\ref{fig:distribution}{\sc\subref{fig:dji_distr}} we have plotted the
bulk of the distributions $p_t(\cdot)$ and $\widehat p_t(\cdot)$ for $t=1$
(daily log-returns) or, more precisely,
the corresponding densities, in the range $[-3\hat s, +3\hat s]$,
where $\hat s \simeq 0.0095$ is the standard deviation of $\widehat p_1(\cdot)$
(i.e., the empirical standard deviation of the daily log returns
evaluated on the DJIA time series).
In Figure~\ref{fig:distribution}{\sc\subref{fig:dji_tails}}
we have plotted the tail of $p_1(\cdot)$,
that is the function $z \mapsto P(X_1 > z) = P(X_1 < -z)$
(note that $X_t \sim -X_t$ for our model) and the right and left
empirical tails $\widehat R(z)$ and $\widehat L(z)$
of $\widehat p_1(\cdot)$, defined for $z\ge 0$ by
 \[
	\widehat{L}(z) := \frac{\#\{1 \le i \le N:\
	x_{i} - x_{i-1} < -z \}}{N}\,, \quad
	\widehat{R}(z) := \frac{\#\{1 \le i \le N:\
	x_{i} - x_{i-1} > z \}}{N} \,,
\] 
in the range $z \in [\hat s, 12\hat s]$.

\begin{figure}
\centering
\subfloat[][\emph{Multiscaling exponent in subperiods of 30 years: simulated data}.]
{\includegraphics[width=.43\columnwidth]{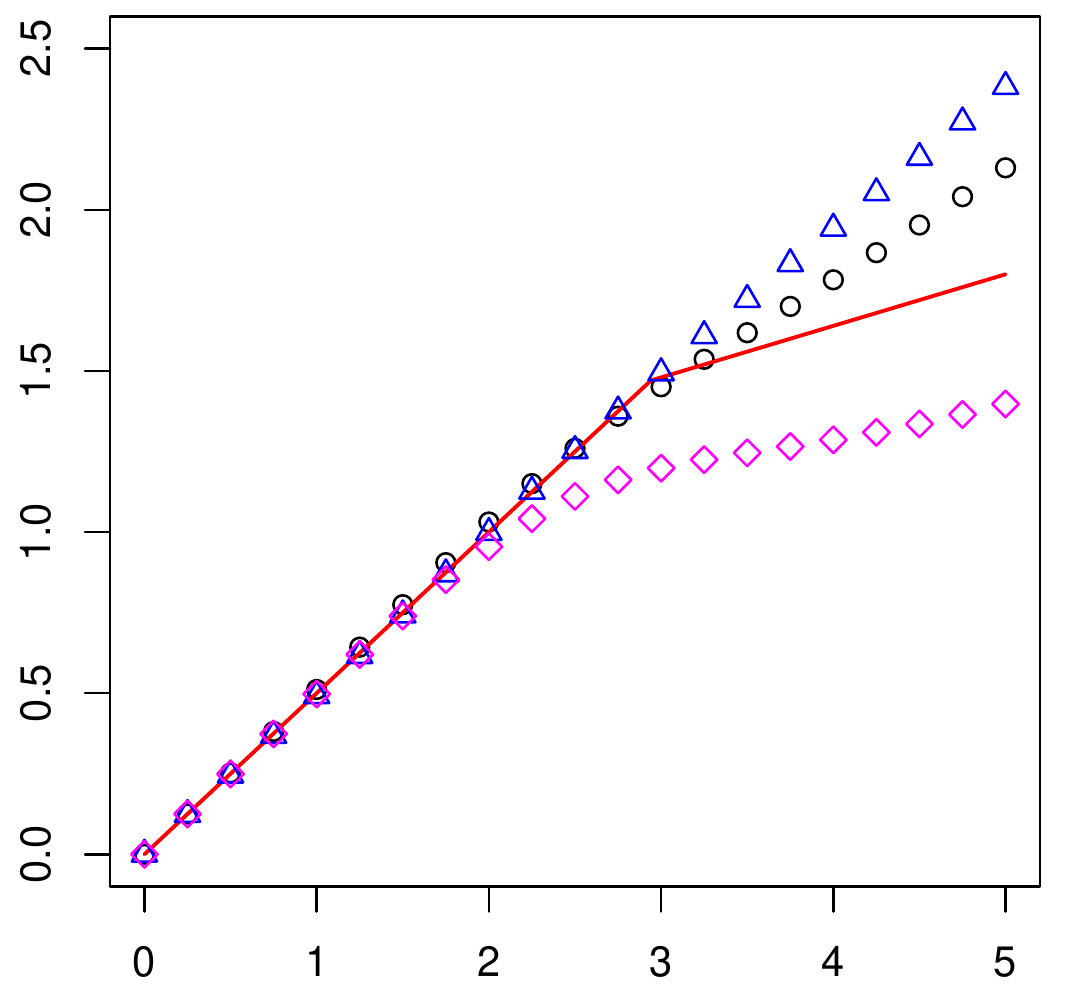}\label{fig:sim_multiscaling-var}}
\qquad \
\subfloat[][\emph{Volatility autocorrelation in subperiods of 30 years: sumilated data}.]
{\includegraphics[width=.43\columnwidth]{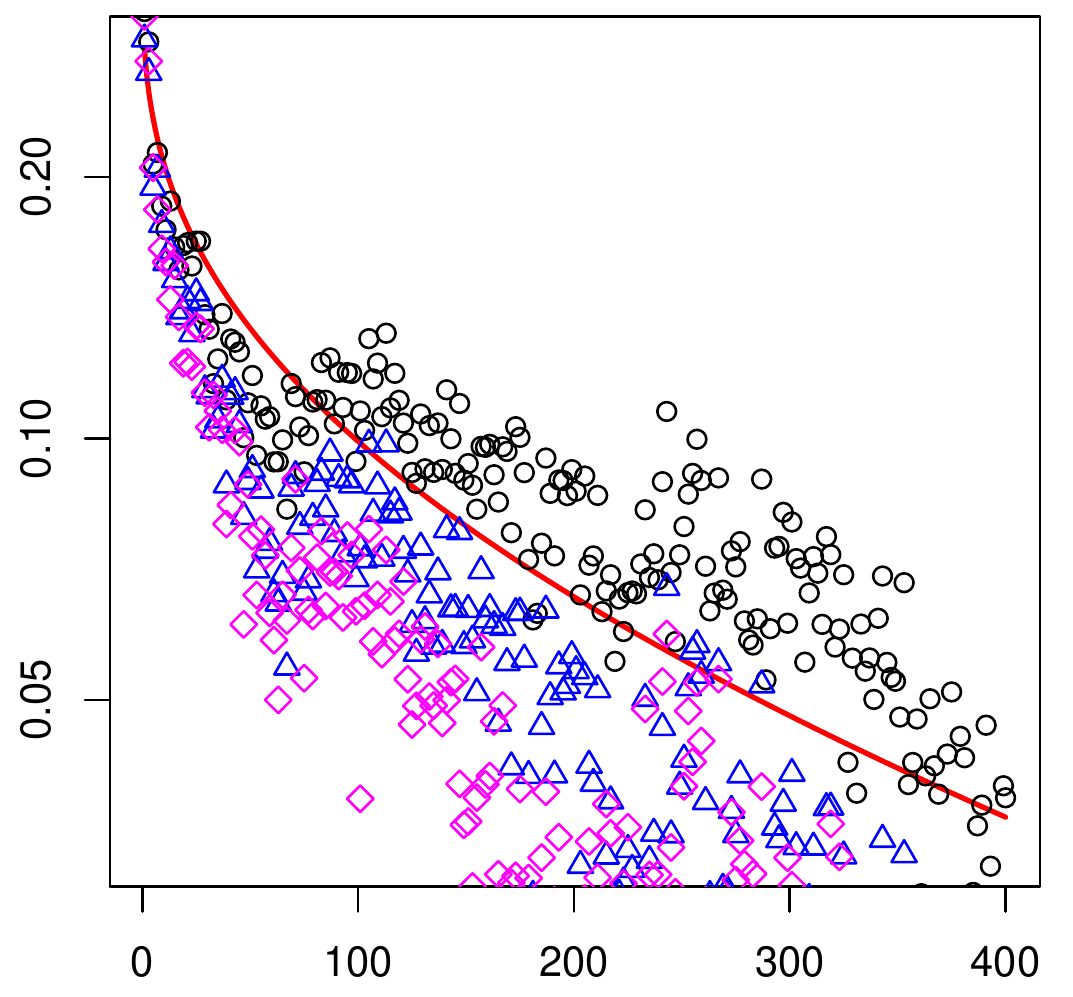}\label{fig:sim_corr400-var}}\\
\smallskip
\subfloat[][\emph{Multiscaling exponent in subperiods of 30 years: DJIA data}.]
{\includegraphics[width=.43\columnwidth]{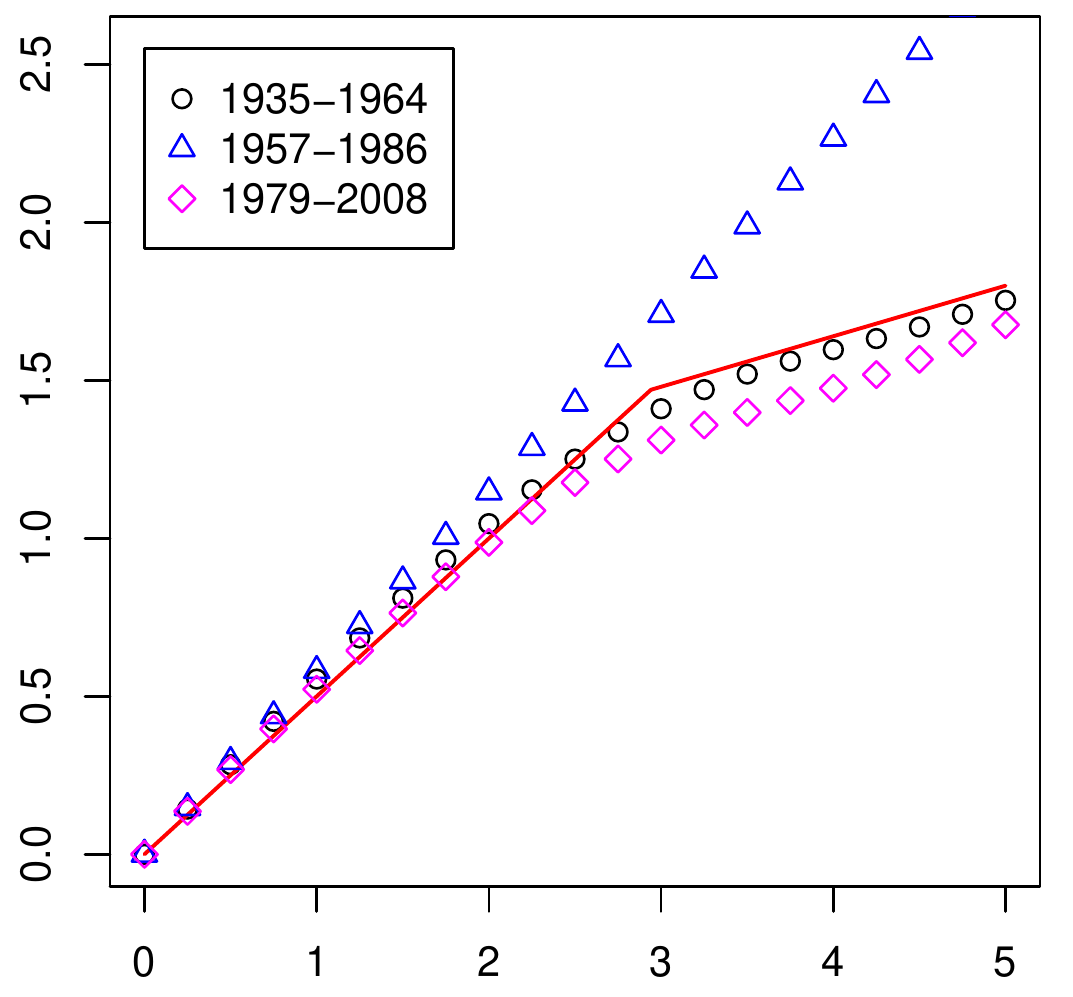}\label{fig:multiscalingvar}}
\qquad \
\subfloat[][\emph{Volatility autocorrelation in subperiods of 30 years: DJIA data}.]
{\includegraphics[width=.43\columnwidth]{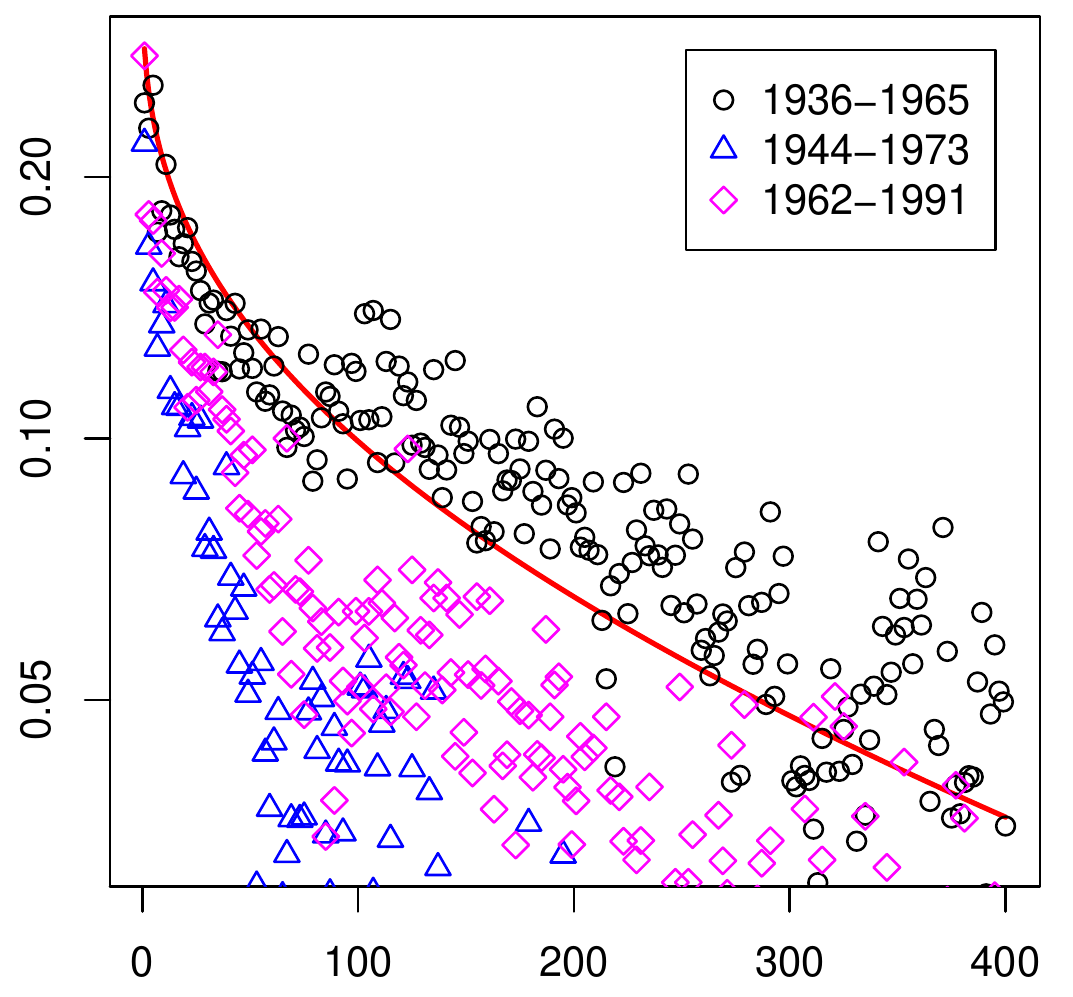}\label{fig:dji_corr400-var}}
\caption{\emph{Variability of estimators in subperiods of 30 years}.\\
Empirical evaluation of the observables $\widehat A(q)$ and $\widehat \rho_1(t)$
in subperiods or 30 years for a 75-years-long time series,
sampled from our model $(X_t)_{t\ge 0}$ ( {\sc\subref{fig:sim_multiscaling-var}} and {\sc\subref{fig:sim_corr400-var}}) 
and from the DJIA time series ({\sc\subref{fig:multiscalingvar}} and {\sc\subref{fig:dji_corr400-var}}) 
}
\label{fig:variability}
\end{figure}

\subsection{Variability of estimators}

\label{sec:variability}

In this paper we have identified relatively rare but dramatic shocks in the volatility 
as the main common source of various 
stylized facts such as multiscaling, autocorrelations and heavy tails. 
As observed in Remark~\ref{rem:robustness}, the expected
number of shocks in a period of 75 years is about 18, which is a rather low number; this means that empirical averages may be not very close to
their ergodic limit or, in different words, estimators should have non-negligible variance. A way to detect this is to simulate data from our model for 75 years, and then compute
estimators using data in different {\em subperiods}, that we have chosen of 30 years.  Figure \ref{fig:variability}{\sc\subref{fig:sim_multiscaling-var}} and
\ref{fig:variability}{\sc\subref{fig:sim_corr400-var}}
show indeed a considerable variability of the values of the estimators for the multiscaling exponent and the volatility autocorrelations, when computed in different subperiods.
We have then repeated the same computations on the DJIA time series,
see Figure~\ref{fig:variability}{\sc\subref{fig:multiscalingvar}} and 
\ref{fig:variability}{\sc\subref{fig:dji_corr400-var}}, and we
have observed a similar variability. We regard this as a significant test for this model.

\begin{remark}
We point out that, among the different quantities that we have
considered, the scaling exponent $\widehat A(q)$ appears to be the most
sensitive. For instance, if instead of the
opening prices one took the closing prices of the DJIA time series (over the same
time period 1935-2009), one would obtain a different (though qualitatively similar)
graph of $\widehat A(q)$.
\end{remark}

\begin{remark}\label{rem:critics}
The multiscaling of empirical moments has been observed in several financial indexes
in \cite{DiM}, where it is claimed that data provide solid arguments \emph{against}
model with linear or piecewise linear scaling exponents.
Note that the theoretical scaling exponent $A(q)$ of our model is indeed
piecewise linear, cf. \eqref{eq:A(q)}. However,
Figure~\ref{fig:variability}{\sc\subref{fig:sim_multiscaling-var}}
shows that the empirical scaling exponent $\widehat{A}(q)$
evaluated on data simulated from our model ``smooths out'' the change of slope,
yielding graphs that are analogous to those obtained for the DJIA time
series, cf. Figure~\ref{fig:variability}{\sc\subref{fig:multiscalingvar}}.
This shows that the objection against models with piecewise linear $A(q)$,
raised in \cite{DiM}, cannot apply to the model we have proposed.
\end{remark}

\medskip
\appendix

\section{Proof of Proposition \ref{prop:etails}}

\label{sec:appetails}

We first need two simple technical lemmas.

\bl{lemma:gausslap}
For $0<q<2$, consider the function $\varphi_q : [0,+\infty) \ra [0,+\infty)$ define by
\[
\varphi_q(\b) := \int_{-\infty}^{+\infty} e^{\b |x|^q - \frac{1}{2} x^2} \, \dd x.
\]
Then there are constants $C_1,C_2 >0$, that depend on $q$, such that for all $\b > 0$
\be{gausslap1}
C_1 e^{C_1 \b^{\frac{2}{2-q}}} \leq \varphi_q(\b) \leq C_2 e^{C_2 \b^{\frac{2}{2-q}}} \,.
\ee
\el

\bpr
We begin by observing that it is enough to establish the bounds in (\ref{gausslap1}) for $\b$ large enough.
Consider the function of positive real variable $f(r) := e^{\b r^q - \frac{1}{2} r^2}$. It is easily checked that $f$ is increasing for $0 \leq r \leq (\b q)^{\frac{1}{2-q}}$. Thus
\[
\varphi_q(\b) \geq \int_{\frac{1}{2} (\b q)^{\frac{1}{2-q}}}^{(\b q)^{\frac{1}{2-q}}} f(r)
\,\dd r \geq \frac{1}{2} (\b q)^{\frac{1}{2-q}} f\left(\frac{1}{2} (\b q)^{\frac{1}{2-q}}\right) = \frac{1}{2} (\b q)^{\frac{1}{2-q}}  \exp\left[c(q) \b^{\frac{2}{2-q}} \right],
\]
with $c(q) := \frac{1}{2^q} q^{\frac{q}{2-q}} - \frac{1}{8}q^{\frac{2}{2-q}} >0$. The lower bound in (\ref{gausslap1}) easily follows for $\b$ large.

For the upper bound, by direct computation one observes that $f(r) \leq e^{- \frac{1}{4} r^2}$ for $r> (4\b)^{\frac{1}{2-q}}$. We have:
\[
\varphi_q(\b) \,\leq\, \int_{|x| \leq (4\b)^{\frac{1}{2-q}}} f(|x|) \, \dd x 
+  \int_{|x| > (4\b)^{\frac{1}{2-q}}} e^{-\frac{1}{4}x^2} \dd x
\,\leq\, 2 (4\b)^{\frac{1}{2-q}} \|f\|_{\infty} + \int_{-\infty}^{+\infty}  e^{-\frac{1}{4}x^2}
\dd x \,.
\]
Since $\|f\|_{\infty} = f((\b q)^{\frac{1}{2-q}}) = \exp\left[C(q)  \b^{\frac{2}{2-q}}\right]$ for a suitable $C(q)$, also the upper bound follows, for $\b$ large.
\epr

\bl{lemma:order}
Let $X_1,X_2,\ldots,X_n$ be independent random variables uniformly distributed in $[0,1]$, and $U_1<U_2< \ldots < U_n$ be the associated order statistics. 
For $n\ge 2$ and $k=2,\ldots,n$, set $\xi_k := U_k - U_{k-1}$.Then, for every $\e>0$
\[
\lim_{n \ra +\infty} P\left( \left| \left\{ k \in \{2, \ldots, n\}:
\ \xi_k > \frac{1}{n^{1+\e}} \right\} \right| \geq n^{1-\e} \right) = 1.
\]
\el

\bpr
This is a consequence of the following stronger result:
for every $x > 0$, as $n\to\infty$ we have the convergence in probability
\begin{equation*}
	\frac 1n \,\left| \left\{ k \in \{2, \ldots, n\}: \ \xi_k > \frac{x}{n} \right\} \right|
	\,\longrightarrow\, e^{-x} \,,
\end{equation*}
see \cite{cf:Weiss} for a proof.
\epr


\begin{proof}[Proof of Proposition \ref{prop:etails}]
Since $X_t = W_{I_t}$ and $\sqrt{I_t} \, W_1$ have the same law,
we can write
\[
E\left[ e^{\g |X_t|^q} \right] = E\left[\exp\left(\g I_t^{q/2} |W_1|^q \right) \right].
\]

We begin with the proof of \eqref{eq:condexpmomB},
hence we work in the regime $q < (1-D)^{-1}$, 
or $q = (1-D)^{-1}$ and $D < \frac 12$; in any case, $q<2$.
We start with the ``$\Leftarrow$'' implication.
Since $I_t$ and $W_1$ are independent, it follows by Lemma~\ref{lemma:gausslap} that
\be{etails1}
E\left[\exp\left(\g I_t^{q/2} |W_1|^q \right) \right] \leq C E\left[ \exp\left( \d I_t^{\frac{q}{2-q}} \right) \right],
\ee
for some $C,\d>0$.
For the moment we work on the event $\{i(t) \ge 1\}$.
It follows by the basic bound \eqref{ub} that
\begin{equation} \label{ubapp}
	I_t \le \sum_{k=0}^{i(t)} \xi_k^{2D} \, \sigma_k^2 \,,
\end{equation}
where we set
\[
\xi_k := \begin{cases}
\tau_1 & \text{for } k=0 \\
\tau_{k+1} - \tau_{k} & \text{for } 1 \leq k \leq i(t)-1 \\
t-\tau_{i(t)} & \text{for } k=i(t)
\end{cases}
\]
Note that $\sum_{k=0}^{i(t)} \xi_k = t$. By applying H\"older inequality to \eqref{ubapp}
with exponents $p = \frac{1}{2D}$, $p' = \frac{1}{1-2D}$, we obtain:
\[
I_t \,\leq\, t^{2D} \left(\sum_{k=0}^{i(t)} \s_{k}^{\frac{2}{1-2D}}\right)^{1-2D} .
\]
%
By assumption $q \le \frac{1}{1-D}$,
which is the same as $(1-2D) \frac{q}{2-q} \le 1$. Thus
\begin{equation}\label{eq:bbbau}
	I_t^{\frac{q}{2-q}} \,\leq\, t^{\frac{2Dq}{2-q}}
	\left(\sum_{k=0}^{i(t)} \s_{k}^{\frac{2}{1-2D}}\right)^{ (1-2D) \frac{q}{2-q} } 
	\,\leq\, t^{\frac{2Dq}{2-q}} \,
	\sum_{k=0}^{i(t)} \s_{k}^{\frac{2q}{2-q}} \,.
\end{equation}
Now observe that if $i(t) = 0$ we have $I_t = \sigma_0^2 [ (t-\tau_0)^{2D} - (-\tau_0)^{2D}]
\le \sigma_0^2 \, t^{2D}$, hence \eqref{eq:bbbau} holds also when $i(t) = 0$.
Therefore, by \eqref{etails1}
\be{asdf}
\begin{split}
	E\left[ e^{\g |X_t|^q} \right] & \,\leq\,
	C E\left[ \exp\left( \d\, t^{\frac{2Dq}{2-q}}
	\sum_{k=0}^{i(t)} \s_k^{\frac{2q}{2-q}} \right) \right]
	\,=\, C \, E\left[ \rho^{i(t)+1} \right] \,,
\end{split}
\ee
where we have set
\[
\rho = \rho_t := E\left[ \exp\left( \delta \, t^{\frac{2Dq}{2-q}}
\, \s_0^{\frac{2q}{2-q}} \right) \right] \,.
\]
Therefore, if $\rho < \infty$, the right hand 
side of \eqref{asdf} is finite, because $i(t) \sim Po(\lambda t)$
has finite exponential moments of all order.
This proves the ``$\Leftarrow$'' implication in \eqref{eq:condexpmomB}.

The ``$\Rightarrow$'' implication in \eqref{eq:condexpmomB} is simpler.
By the lower bound in Lemma \ref{lemma:gausslap} we have
\be{etails4}
E\left[ e^{\g |X_t|^q} \right]  = E\left[\exp\left(\g I_t^{q/2} |W_1|^q \right) \right] \geq C E\left[ \exp\left( \d I_t^{\frac{q}{2-q}} \right) \right],
\ee
for suitable $C,\d>0$. We note that
\be{etails5}
\begin{split}
	E\left[ \exp\left( \d I_t^{\frac{q}{2-q}} \right) \right] 
	& \geq  E\left[ \exp\left( \d I_t^{\frac{q}{2-q}} \right) 
	\ind_{\{i(t) = 0\}}\right]  \\
	& = E\left[ \exp\left( \d \left[(t-\tau_0)^{2D} - (-\tau_0)^{2D}\right]^{\frac{q}{2-q}}
	\, \s_0^{\frac{2q}{2-q}} \right) \right] P( i(t) = 0 ) \,.
\end{split}
\ee
Under the condition
\[
E\left[\exp\left( \a \s^{\frac{2q}{2-q}} \right) \right] = +\infty \ \forall \a>0,
\]
the last expectation in \eqref{etails5} is infinite, since
$\left[(t-\tau_0)^{2D} - (-\tau_0)^{2D}\right] > 0$ almost surely
and is independent of $\s_0$. Looking back
at \eqref{etails4}, we have proved the
``$\Rightarrow$'' implication in \eqref{eq:condexpmomB}.

\smallskip

Next we prove \eqref{eq:condexpmomA}, hence we assume
that $q > (1-D)^{-1}$.
Consider first the case $q < 2$ (which may happen only for $D < \frac 12$).
By \eqref{etails4}
\begin{equation*}
	E\left[ e^{\g |X_t|^q} \right]  
	\geq C E\left[ \exp\left( \d I_t^{\frac{q}{2-q}} \right) \right] \,.
\end{equation*}
We note that, by the definition \eqref{eq:it} of $I_t$, we can write
\be{etails6}
I_t \geq \sum_{k=2}^{i(t)} \s_{k-1}^2 (\tau_k - \tau_{k-1})^{2D} \,,
\ee
where we agree that the sum is zero if $i(t) < 2$.
For $n \geq 0$, we let $P_n$ to denote the conditional probability
$P(\,\cdot\,|i(t) = n)$ and $E_n$ the corresponding expectation.
Note that, under $P_n$, the random variables $ (\tau_k - \tau_{k-1})_{k=2}^n$ have 
the same law of the random variables $(\xi_k)_{k=2}^n$ in Lemma~\ref{lemma:order},
for $n \geq 2$. Consider the following events:
\[
	A_n := \{ \s_k^2 \geq a , \ \forall k=2,\ldots,n\}\,,
	\qquad
	B_n := \left\{
	\left| \left\{ k= 2,\ldots,n: \ \xi_k > \frac{1}{n^{1+\e}} \right\}
	\right| \geq n^{1-\e}  \right\},
\]
where $a>0$ is such that $\nu([a,+\infty)) =: \rho > 0$ and $\e>0$ will be chosen later. 
Note that $P_n(A_n) = \rho^{n-1}$ while $P_n(B_n) \to 1$ as $n \ra +\infty$,
by Lemma~\ref{lemma:order}. In particular, there is $c>0$ such that
$P_n(B_n) \geq c$ for every $n$.
Plainly, $A_n$ and $B_n$ are independent under $P_n$. We have
\begin{multline} \label{etails7}
\psi(n) :=  E_n \left[ \exp\left( \d I_t^{\frac{q}{2-q}} \right) \right]  \geq  E_n \left[ \exp\left( \d I_t^{\frac{q}{2-q}} \right) \ind_{A_n \cap B_n} \right] \\ \geq 
c \, \rho^{n-1} \exp\left[ \d a^{\frac{q}{2-q}} \left( \frac{1}{n^{1+\e}} \right)^{2D \frac{q}{2-q}} n^{(1-\e) \frac{q}{2-q}} \right] \\ = 
c \, \rho^{n-1}\exp\left[ \d a^{\frac{q}{2-q}} n^{(1-2D-\e (1+2D)) \frac{q}{2-q}} \right]
\end{multline}
Note that $q > \frac{1}{1-D}$ is equivalent to
$(1-2D) \frac{q}{2-q} > 1$, therefore $\e$ can be chosen small enough so that
$ b:= (1-2D-\e (1+2D)) \frac{q}{2-q}>1$.
It then follows by \eqref{etails7} that $\psi(n) \ge d \, \exp(d \, n^{b})$
for every $n\in\N$, for a suitable $d>0$. Therefore
\[
E\left[ \exp\left( \d I_t^{\frac{q}{2-q}} \right) \right] = E[\psi(i(t))] = +\infty \,,
\]
because $i(t) \sim Po(\lambda t)$ and hence $E[\exp(d \, i(t)^b)] = \infty$
for all $d > 0$ and $b > 1$.

Next we consider the case $q \geq 2$. Note that
\begin{equation} \label{eq:asdf}
	E\left[ e^{\g |X_t|^q} \right]
	= E\left[\exp\left(\g I_t^{q/2} |W_1|^q \right) \right] \,,
\end{equation}
hence if $q>2$ we have $E\left[ e^{\g |X_t|^q} \right] = \infty$,
because $E[\exp(c |W_1|^q)] = \infty$
for every $c > 0$, $I_t > 0$ almost surely and $I_t$ is independent of $W_1$.
On the other hand, if $q=2$ we must have $D < \frac 12$ (recall that we
are in the regime $q > (1-D)^{-1}$) and
the steps leading to (\ref{etails7}) have shown that in this case $I_t$ is unbounded.
It then follows again from \eqref{eq:asdf}
that $E[ e^{\g |X_t|^2} ] = \infty$.
\end{proof}


\medskip
\section{The model of Baldovin and Stella}
\label{sec:BalSte}

Let us briefly discuss the model proposed by F. Baldovin and A. Stella \cite{BV1,BV2},
motivated by renormalization group arguments from statistical physics.
They first introduce a process $(Y_t)_{t \geq 0}$ 
which satisfies the scaling relation \eqref{eq2} 
for a given function $g$, that is assumed to be even, so that
its Fourier transform $\hat g(u) := \int_\R e^{i u x} g(x) \dd x$ is real
(and even).
The process $(Y_t)_{t \geq 0}$ is defined by
specifying its finite dimensional laws:
for $t_1 < t_2 < \cdots < t_n$ the joint density of
$Y_{t_1}, Y_{t_2}, \ldots , Y_{t_n}$ is given by
\be{eq3}
p(x_1,t_1;x_2,t_2; \ldots ; x_n,t_n) = h\left( \frac{x_1}{\sqrt{t_1}}, \frac{x_2 - x_1}{\sqrt{t_2 - t_1}}, \ldots , \frac{x_n - x_{n-1}}{\sqrt{t_n - t_{n-1}}} \right) ,
\ee
where $h$ is the function whose Fourier transform $\hat h$ is given by
\be{eq4}
\hat{h}(u_1,u_2,\ldots,u_n) := \hat{g} \Big(
\sqrt{ u_1^2 + \ldots + u_n^2} \Big) .
\ee
Note that if $g$ is the standard Gaussian density,
then $(Y_t)_{t \geq 0}$ is the ordinary Brownian motion.
For a non Gaussian $g$, the expression in \eqref{eq4} is not necessarily
the Fourier transform of a probability on $\R^n$, so that some care is needed
(we come back to this point in a moment).
However, it is clear from \eqref{eq3} that the increments of the 
process $(Y_t)_{t \geq  0}$ corresponding to time intervals
of the same length (that is, for fixed $t_{i+1}-t_i$)
have a \emph{permutation invariant distribution} and therefore
cannot exhibit any decay of correlations.

For this reason, Baldovin and Stella introduce what is
probably the most interesting ingredient of their construction, namely
a special form of time-inhomogeneity.
They define it in terms of finite dimensional distributions, 
bur it is simpler to give a pathwise construction:
given a sequence of (possibly random) times $0 < \tau_1 < \tau_2 < \cdots < \tau_n \uparrow +\infty$
and a fixed $0 < D \le 1/2$, they introduce a new process $(X_t)_{t\ge 0}$
defined by
\begin{equation} \label{eq56}
	X_t := Y_{t^{2D}}
	\qquad \text{for } t \in [0, \tau_1),
\end{equation}
and more generally
\begin{equation} \label{eq6}
	X_t := Y_{(t-\tau_n)^{2D} + \sum_{k=1}^{n} 
	(\tau_k - \tau_{k-1})^{2D}}
	\qquad \text{for } t \in [\tau_n,\tau_{n+1}) \,.
\end{equation}
For $D= 1/2$ we have clearly $X_t \equiv Y_t$,
while for $D < 1/2$ the process $(X_t)_{t\ge 0}$ is obtained from $(Y_t)_{t\ge 0}$ 
by a nonlinear time-change, that is ``refreshed'' at each time $\tau_n$.
This transformation has the effect of amplifying 
the increments of the process for $t$ immediately after the
times $(\tau_n)_{n\ge 1}$, while the increments tend to become small for larger $t$.

%

\smallskip

Let us shed some light into the implicit relations
\eqref{eq3}--\eqref{eq4}.
If a stochastic process $(Y_t)_{t \geq 0}$ 
is to satisfy these relations, it must necessarily
have \emph{exchangeable increments}: by this we mean
(cf. \cite[p.1210]{Freedman}) that,
setting $\Delta Y_{(a,b)} := Y_{b} - Y_a$ for short,
the distribution of the random vector $(\Delta Y_{I_1 + y_1},\,
\ldots,\, \Delta Y_{I_n + y_n})$ --- where the $I_j$'s are intervals
and $y_j$'s real numbers --- does not depend on $y_1, \ldots, y_n$, as long as the intervals
$y_1 + I_1$, \ldots, $y_n + I_n$ are disjoint.
If we make the (very mild) assumption that $(Y_t)_{t \geq 0}$
has no fixed point of discontinuity, 
then a continuous-time version of the celebrated de Finetti's theorem
ensures that $(Y_t)_{t \geq 0}$ is a mixture of L\'evy processes,
cf. Theorem~3 in~\cite{Freedman} (cf. also~\cite{Ac:Lu}).
Actually, more can be said: since by \eqref{eq2} the distribution of
the increments of $(Y_t)_{t \geq 0}$ is \emph{isotropic},
i.e., it has spherical symmetry in $\R^n$,
by Theorem~4 in~\cite{Freedman}
the process $(Y_t)_{t \geq 0}$ is necessarily \emph{a mixture of Brownian motions}.
This means that we have the following representation:
\be{eq7}
Y_t = \s\, W_t,
\ee
where $(W_t)_{t\ge 0}$ is a standard Brownian motion and $\s$ is an independent real
random variable (a random, but time-independent, volatility).
Viceversa, if a process $(Y_t)_{t\ge 0}$ satisfies \eqref{eq7},
then, denoting by $\nu$ the law of $\sigma$, it is easy to check that
relations \eqref{eq3}--\eqref{eq4} hold with
\be{eq8}
	g(x) = \int_\R \frac{1}{\sqrt{2 \pi \s}}
	\, e^{-\frac{x^2}{2 \s^2}} \,\nu(\dd\s) \,,
\ee
or, equivalently,
\[
\hat{g}(u) = \int_\R e^{-\frac{\s^2 u^2}{2 }} \,  \nu(\dd\s) \,.
\]
This shows that the
functions $g$ for which \eqref{eq3}--\eqref{eq4} provide a consistent 
family of finite dimensional distributions are exactly those
that may be expressed as in \eqref{eq8} for some probability $\nu$ on $(0,+\infty)$.

Note that a path of \eqref{eq7} is obtained by sampling independently $\s$ from $\nu$ 
and $(W_t)_{t\ge 0}$ from the Wiener measure, hence this path 
{\em cannot} be distinguished  from the path of a Brownian motion with 
{\em constant} volatility. In particular, the (possible) correlation of the increments
of the process $(Y_t)_{t\ge 0}$ cannot be detected empirically, and the same observation applies
to the time-inhomogeneous process $(X_t)_{t\ge 0}$ obtained by 
$(Y_t)_{t\ge 0}$ through \eqref{eq56}--\eqref{eq6}.
In other words, the processes obtained through this construction have non ergodic increments.

Nevertheless, Baldovin and Stella claim to
measure nonzero correlations from their samples:
after estimating the function $g$ and the parameters $t_0$ and $D$ on the DJIA
time series, their simulated trajectories show a good 
agreement with the clustering of volatility,
as well as with the basic scaling \eqref{eq2} and the multiscaling of moments.
The explanation of this apparent contradiction is that Baldovin and Stella
do not simulate the process $(X_t)_{t\ge 0}$ defined through the above construction,
but rather an autoregressive approximation of it.
In fact, besides making a periodic choice of the times $\tau_n := n t_0$,
they fix a small time step $\delta$ and a natural number $N$ and
they first simulate $x_{\delta}$, $x_{2\delta}$, \ldots, $x_{N \delta}$
according to the true distribution of $(X_{\delta}, X_{2\delta}, \ldots, X_{N \delta})$.
Then they compute the conditional distribution of $X_{(N+1)\delta}$ given
$X_{2\delta} = x_{2\delta}$, $X_{3\delta} = x_{3\delta}$, \ldots, 
$X_{N \delta} = x_{N\delta}$ --- thus \emph{neglecting} $x_{\delta}$ ---
and sample $x_{(N+1)\delta}$ from this distribution. 
Similarly, $x_{(N+2)\delta}$ is sampled from the conditional distribution of 
$X_{(N+2)\delta}$ given $X_{3\delta} = x_{3\delta}$, \ldots,
$X_{N\delta} = x_{N\delta}$,  $X_{(N+1) \delta} = x_{(N+1)\delta}$,
neglecting both $x_\delta$ and $x_{2\delta}$, and so on.
It is plausible that such an autoregressive procedure may produce an ergodic
process.


\medskip

\medskip
\section*{Acknowledgements}

We thank Fulvio Baldovin, Massimiliano Caporin, Wolfgang Runggaldier
and Attilio Stella for fruitful discussions.
We are very grateful to the anonymous referee for several important
remarks and suggestions.


\end{document}